\documentclass[aps,twocolumn,groupedaddress,showpacs]{revtex4}

\begin{document}
\bibliographystyle{apsrev}

\title{
Quantum order from string-net condensations
and origin of light and massless fermions
}

\author{Xiao-Gang Wen}
\homepage{http://dao.mit.edu/~wen}
\affiliation{Department of Physics, Massachusetts Institute of Technology,
Cambridge, Massachusetts 02139}

\date{Dec. 2002}

\begin{abstract}
Recently, it was pointed out that quantum orders and the associated
projective symmetry groups can produce and protect massless gauge bosons and
massless fermions in local bosonic models. In this paper, we demonstrate that
a state with such kind of quantum orders can be viewed as a string-net condensed
state.  The emerging gauge bosons and fermions in local bosonic models can be
regarded as a direct consequence of string-net condensation.
The gauge bosons are fluctuations of large 
closed string-nets which are condensed in the ground
state.  The ends of open strings (or nodes of open string-nets)
are the charged particles of the
corresponding gauge field.  For certain types of strings, the nodes of
string-nets
can even be fermions.  According to the string-net picture, 
fermions always carry gauge charges. This suggests the existence of
a new discrete gauge field that couples to neutrinos and neutrons.  We also
discuss how chiral symmetry that protects massless Dirac fermions can emerge
from the projective symmetry of quantum order.
\end{abstract}

\maketitle

\tableofcontents

\section{Introduction}

\subsection{Fundamental questions about light and fermions}
\label{fq}

We have known light and fermions for many years. But we still cannot give a
satisfactory answer to the following fundamental questions: What are light and
fermions?  Where light and fermions come from?  Why light and fermions exist?
At moment, the standard answers to the above fundamental questions appear to
be ``light is the particle described by a gauge field'' and ``fermions are the
particles described by anti-commuting fields''.  Here, we like to argue that
there is another possible answer to the above questions: our vacuum is filled
with  string-like objects that form network of arbitrary sizes and those
string-nets form a quantum condensed state.  According to the string-net
picture, the light (and other gauge bosons) is a vibration of the condensed
string-nets and fermions are ends of strings (or nodes of string-nets).  The
string-net condensation provides a unified origin of light and
fermions.\footnote{Here, by ``string-net condensation'' we mean condensation
of nets of string-like objects of arbitrary sizes.}

Before discussing the above fundamental questions in more detail, we would
like to clarify what do we mean by ``light exists'' and ``fermions exist''. We
know that there is a natural mass scale in physics -- the Planck mass. Planck
mass is so large that any observed particle have a mass at least factor
$10^{16}$ smaller than the Planck mass.  So all the observed particles can be
treated as massless when compared with Planck mass. When we ask why some
particles exist, we really ask why those particles are massless (or nearly
massless when compared with Planck mass).  So the real issue is to understand
what makes certain excitations (such as light and fermions) massless.
We have known that symmetry breaking is a way to get gapless bosonic 
excitations. We will see that string-net condensation is another way to get 
gapless excitations. However, string-net condensations can generate
massless gauge bosons and massless fermions.

Second, we would like to clarify what do we mean by ``origin of light and
fermions''. We know that everything has to come from something. So when we ask
``where light and fermions come from'', we have assumed that there are some
things simpler and more fundamental than light and fermions.  In the section
\ref{LBM}, we define local bosonic models which are simpler than models with
gauge fields coupled to fermions.  We will regard local bosonic models as more
fundamental (the locality principle).  We will show that light and fermions
can emerge from a local bosonic model if the model contains a condensation of
nets of string-like object in its ground state.

After the above two clarifications, we can state more precisely the meaning
of ``string-net condensation provides another possible answer to the
fundamental questions about light and fermions''. When we say gauge bosons and
fermions originate from string-net condensation, we really mean that (nearly)
\emph{massless} gauge bosons and fermions originate from string-net
condensation in a \emph{local bosonic model}.

\subsection{Gapless phonon and symmetry breaking orders}

Before considering the origin of massless photon and massless fermions, let us
consider a simpler massless (or gapless) excitation -- phonon. We can ask
three similar questions about phonon: What is phonon? Where phonon comes from?
Why phonon exists?  We know that those are scientific questions and we know
their answers.  Phonon is a vibration of a crystal. 
Phonon comes from a spontaneous translation symmetry breaking. Phonon exists
because the translation-symmetry-breaking phase actually exists in nature.  In
particular, the gaplessness of phonon is directly originated from and
protected by the spontaneous translation symmetry breaking.\cite{N6080,G6154}
Many other gapless excitations, such as spin wave, superfluid mode \etc, also
come from condensation of point-like objects that break certain symmetries.

It is quite interesting to see that our understanding of a gapless excitation
- phonon - is rooted in our understanding of phases of matter.  According to
  Landau's theory,\cite{L3726} phases of matter are different because they
have different broken symmetries.  The symmetry description of phases is very
powerful. It allows us to classify all possible crystals. It also provides the
origin for gapless phonons and many other gapless excitations. Until a few
years ago, it was believed that the condensations of point-like objects, and
the related symmetry breaking and order parameters, can describe all the
orders (or phases) in nature.

\subsection{The existence of light and fermions implies the existence of new
orders}

Knowing light as a massless excitation, one may wonder maybe light, just like
phonon, is also a Nambu-Goldstone mode from a broken symmetry. However,
experiments tell us that a $U(1)$ gauge boson, such as light, is really
different from a Nambu-Goldstone mode in 3+1 dimensions.  Therefore it is
impossible to use Landau's symmetry breaking theory and condensation of
point-like objects to understand the origin and the masslessness of light.
Also, Nambu-Goldstone modes are always bosonic, thus it is impossible to use
symmetry breaking to understand the origin and the (nearly) masslessness of
fermions.  It seems that there does not exist any order that can give rise to
massless light and massless fermions.  Because of this, we put light and
electron into a different category than phonon. We regarded them as elementary
and introduced them by hand into our theory of nature.

However, if we believe light and electrons, just like phonon, exist for a
reason, then such a reason must be a certain order in our vacuum that protect
the masslessness of light and electron.  (Here we have assumed that light and
electron are not something that we place in an empty vacuum.  Our vacuum is
more like an ``ocean'' which is not empty. Light and electron are collective
excitations that correspond to certain patterns of ``water'' motion.) Now the
question is that what kind of order can give rise to light and electron, and
protect their masslessness.  

If we really believe in the equality between light, electron and phonon, then
the very existence of light and fermions indicates that our understanding of
states of matter is incomplete. We should deepen and expand our understanding
of the states of matter. There should be new states of matter that contain new
kind of orders.  The new orders will  produce light and electron, and protect
their masslessness.

\subsection{Topological order and quantum order}

After the discovery of fractional quantum Hall (FQH)
effect,\cite{TSG8259,L8395} it became clear that the Landau's symmetry
breaking theory cannot describe different FQH states, since those states all
have the \emph{same} symmetry. It was proposed that FQH states contain a new
kind of order - topological order.\cite{Wtoprev} Topological order is new
because it cannot be described by symmetry breaking, long range correlation,
and local order parameters.  Non of the usual tools that we used to
characterize phases applies to topological order. Despite of this, topological
order is not an empty concept.  Topological order can be characterized by a new
set of tools, such as the number of degenerate ground states, quasiparticle
statistics, and edge states.  It was shown that the ground state degeneracy of
a topological ordered state is a universal property since the degeneracy is
robust against any perturbations.\cite{WNtop} Such a topological degeneracy
demonstrates the existence of topological order. It can also be used to perform
fault tolerant quantum computations.\cite{K9721}

\begin{figure}
\centerline{
\includegraphics[width=3.6in]{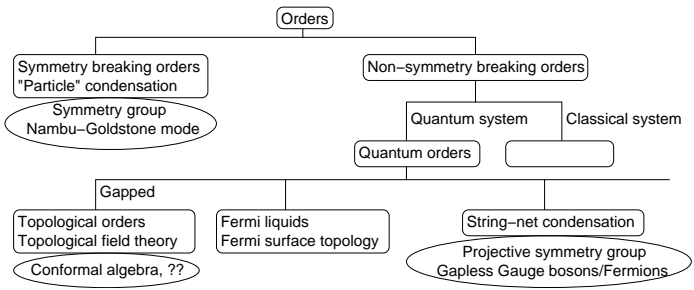}
}
\caption{
A classification of different orders in matter.
(We view our vacuum as one kind of matter.)
}
\label{clorder}
\end{figure}

Recently, the concept of topological order was generalized to quantum
order.\cite{Wqoslpub,Wqogen} Quantum order is used to describe new kinds of
orders in gapless quantum states. One way to understand quantum order is to
see how it fits into a general classification scheme of orders (see Fig.
\ref{clorder}).  First, different orders can be divided into two classes:
symmetry breaking orders and non-symmetry breaking orders. The symmetry
breaking orders can be described by a local order parameter and can be said to
contain a condensation of point-like objects. All the symmetry breaking orders
can be understood in terms of Landau's symmetry breaking theory. The
non-symmetry breaking orders cannot be described by symmetry breaking, neither
by the related local order parameters and long range correlations. Thus they
are a new kind of orders. If a quantum system (a state at zero temperature)
contains a non-symmetry breaking order, then the system is said to contain a
non-trivial quantum order. We see that a quantum order is simply a
non-symmetry breaking order in a quantum system.

Quantum order can be further divided into many subclasses. If a quantum state
is gapped, then the corresponding quantum order will be called topological
order. The low energy effective theory of a topological ordered state will be
a topological field theory.\cite{W8951} The second class of quantum orders
appear in Fermi liquids (or free fermion systems). The different quantum
orders in Fermi liquids are classified by the Fermi surface
topology.\cite{L6030,Wqogen}

\subsection{The quantum orders from string-net condensations}

In this paper, we will concentrate on the third class of quantum orders -- the
quantum orders from condensation of nets of strings, or simply, string-net 
condensation.\cite{Walight,LW0360} This class of
quantum orders shares some similarities with the symmetry breaking orders of
``particle'' condensation. We know that different symmetry breaking orders can
be classified by symmetry groups. Using group theory, we can classify all the
230 crystal orders in three dimensions. The symmetry also produces and
protects gapless Nambu-Goldstone bosons. Similarly, as we will see later in
this paper, different string-net condensations (and the corresponding quantum
orders) can be classified by a mathematical object called projective symmetry
group (PSG).\cite{Wqoslpub,Wqogen} Using PSG, we can classify over 100
different 2D spin liquids that all have the same symmetry.\cite{Wqoslpub} Just
like symmetry group, PSG can also produce and protect gapless excitations.
However, unlike symmetry group, PSG produces and protects gapless gauge bosons
and gapless fermions.\cite{Wqoslpub,Wlight,WZqoind} Because of this, we can
say light and massless fermions can have a unified origin. They can come from
string-net condensations.

We used to believe that to have light and fermions in our theory, we have to
introduce by hand a fundamental $U(1)$ gauge field and anti-commuting fermion
fields, since at that time we did not know any collective modes that behave
like gauge bosons and fermions. Now, we know that
gauge bosons and fermions appear commonly and naturally in quantum ordered
states, as fluctuations of condensed string-nets and ends of open strings. 
This raises an issue: do light and fermions come from a
fundamental $U(1)$ gauge field and anti-commuting fields as in the 123
standard model or do they come from a particular quantum order in our vacuum?
Clearly it is more natural to assume light and fermions come from a quantum
order in our vacuum.  From the connection between string-net condensation, quantum
order, and massless gauge/fermion excitations, it is very tempting to 
propose the following answers to the fundamental questions about
light and (nearly) massless fermions:\vskip 1mm
\noindent
\fbox{\parbox{3.3in}{
{\bf What are light and fermions?} \\
Light is a fluctuation of condensed string-nets of arbitrary sizes.
Fermions are ends of open strings.
}}\vskip 2mm
\noindent
\fbox{\parbox{3.3in}{
{\bf Where light and (nearly) massless fermions come from?}  \\
Light and the fermions come from the collective motions of nets of
string-like objects that fill our vacuum.
}}
\vskip 2mm \noindent
\fbox{\parbox{3.3in}{
{\bf Why light and (nearly) massless fermions exist?}   \\
Light and the fermions exist because our vacuum 
chooses to have a string-net condensation.
}}
\vskip 2mm

Had our vacuum chosen to have a ``particle'' condensation, there would be only
Nambu-Goldstone bosons at low energies. Such a universe would be very boring.
String-net condensation and the resulting light and (nearly) massless fermions
provide a much more interesting universe, at least interesting enough to
support intelligent life to study the origin of light and massless fermions.

The string-net picture of fermions explains why there is always an even number
of fermions in our universe.  The string-net picture for gauge bosons and
fermions also has an experimental prediction: all fermions must carry certain
gauge charges.\cite{LW0360}  At first sight, this prediction appears to
contradict with the known experimental fact that neutron carry no gauge
charges. Thus one may think the string-net picture of gauge bosons and fermions
has already been falsified by experiments.  Here we would like to point out
that the string-net picture of gauge bosons and fermions can still be correct if
we assume the existence of a new discrete gauge field, such as a $Z_2$ gauge
field, in our universe. In this case, neutrons and neutrinos carry a non-zero
charge of the discrete gauge field. Therefore, the string-net picture of gauge
bosons and fermions predict the existence of discrete gauge excitations (such
as gauge flux lines) in our universe.

We would like to remark that, despite the similarity, the above string-net
picture of gauge bosons and fermions is different from the picture of standard
superstring theory. In standard superstring theory, closed strings correspond
to gravitons, and open string correspond to gauge bosons. All the elementary
particles correspond to different vibration modes of small strings in the
superstring theory.  Also, the fermions in the standard superstring theory
come from the fermion fields on the world sheet.  In our string-net picture,
the vacuum is filled with large nets of strings. The massless gauge bosons
correspond to the fluctuations of large closed string-nets (\ie nets of closed
strings) and fermions  correspond to the ends of open strings in string nets.
Anti-commuting fields are not needed to produce (nearly) massless fermions.
Massless fermions appear as low energy collective modes in a purely bosonic
system.  

The string-net picture for gauge theories have a long history.  The
closed-string description of gauge fluctuations is intimately related to the
Wilson loop in gauge theory.\cite{W7159,W7445,K7959} The relation between
dynamical gauge theory and a dynamical Wilson-loop theory was suggested in
\Ref{P7947,P981}.  \Ref{KS7595} studied the Hamiltonian of a non-local model -
lattice gauge theory. It was found that the lattice gauge theory contains a
string-net structure and the gauge charges can be viewed as ends of strings.  In
\Ref{S8053,ID89} various duality relations between lattice gauge theories and
theories of extended objects were reviewed. In particular, some statistical
lattice gauge models were found to be dual to certain statistical membrane
models.\cite{BMK7793} This duality relation is directly connected to the
relation between gauge theory and closed-string-net theory\cite{Walight} 
in quantum models.

Emerging fermions from local bosonic models also have a complicated history.
The first examples of emerging fermions/anyons were the fractional quantum
Hall states,\cite{TSG8259,L8395} where fermionic/anyonic excitations were
obtained theoretically from interacting bosons in magnetic
field.\cite{ASW8422} In 1987, fermion fields and gauge fields were introduced
to express the spin-1/2 Hamiltonian in the slave-boson
approach.\cite{BZA8773,BA8880} However, writing a bosonic Hamiltonian in terms
of fermion fields does not imply the appearance of well defined fermionic
quasiparticles.
Emerging fermionic excitations can appear only in deconfined phases of the
gauge field.  \Ref{WWZcsp,KW8983,RS9173,Wsrvb} constructed several
deconfined phases where the fermion fields do describe well defined
quasiparticles. However, depending on the property of deconfined phases, those
quasiparticles may carry fractional statistics (for the chiral spin
states)\cite{KL8795,WWZcsp,KW8983} or Fermi statistics (for the $Z_2$
deconfined states).\cite{RS9173,Wsrvb} 

Also in 1987, in a study of resonating-valence-bond (RVB) states, emerging
fermions (the spinons) were proposed in a nearest neighbor dimer model on
square lattice.\cite{KRS8765,RK8876,RC8933} But, according to the
deconfinement picture, the results in \Ref{KRS8765,RK8876} are valid only when
the ground state of the dimer model is in the $Z_2$ deconfined phase. It
appears that the dimer liquid on square lattice with only nearest neighbor
dimers is not a deconfined state,\cite{RK8876,RC8933} and thus it is not clear
if the nearest neighbor dimer model on square lattice\cite{RK8876} has the
fermionic quasiparticles or not.\cite{RC8933}  However, on triangular lattice,
the dimer liquid is indeed a $Z_2$ deconfined state.\cite{MS0181} Therefore,
the results in \Ref{KRS8765,RK8876} are valid for the triangular-lattice dimer
model and fermionic quasiparticles do emerge in a dimer liquid on triangular
lattice.  

All the above models with emerging fermions are 2+1D models, where the
emerging fermions can be understood from binding flux to a charged
particle.\cite{ASW8422} Recently, it was pointed out in \Ref{LW0360} that the
key to emerging fermions is a string structure. Fermions can generally appear
as ends of open strings. The string picture allows a construction
of a 3+1D local bosonic model that has emerging fermions.

Comparing with those previous results, the new features discussed in this
paper are: (A) \emph{Massless} gauge bosons and fermions can
emerge from \emph{local bosonic models} as a result of string-net condensation.
(B) Massless fermions are protected by the string-net condensation (and the
associated PSG).  (C) String-net condensed states represent a new kind of phases
which cannot be described Landau's symmetry breaking theory. Different
string-net
condensed states are characterized by different PSG's. 
(D) QED and QCD can emerge from a local bosonic model on cubic lattice.
The effective QED and QCD has $4N$ families of leptons and quarks. Each family
has one lepton and two flavors of quarks.

The bottom line is that, within local bosonic models, massless fermions do not
just emerge by themselves. Emerging massless fermions, emerging massless gauge
bosons, string-net condensations, and PSG are intimately related. They are just
different sides of same coin
- quantum order.

According to the picture of quantum order, elementary particles (such as
photon and electron) may not be elementary after all. They may be collective
excitations of a local bosonic system below Planck scale.  Since we cannot do
experiments close to Planck scale, it is hard to determine if photon and
electron are elementary particles or not. In this paper, we would like to show
that the string-net picture of light and fermions is at least self consistent by
studying some concrete local boson models which produce massless gauge bosons
and massless fermions through string-net condensations.  The local boson models
studied here are just a few examples among a long list of local boson
models\cite{DDL7863,W7985,KL8795,BA8880,AM8874,KL8842,RK8876,WWZcsp,RS9173,%
Wsrvb,K9721,MS0181,SP0258,BFG0212,MS0204,IFI0203,Wqoexct,M0284} that contain
emerging fermions and gauge fields.

Here we would like to stress that the string-net picture for the actual gauge
bosons and fermions in our universe is only a proposal at moment.  Although
string-net condensation can produce and protect massless photons, gluons,
quarks, and other charged leptons, we do not know at moment if string-net
condensations can produce neutrinos which are chiral fermions, and the
weak-interaction $SU(2)$ gauge field which couples chirally to the quarks and
the leptons.  Also, we do not know if string-net condensation can produce an
odd number of families of quarks and leptons. The QED and QCD produced by the
known string-net condensations all contain an even number of families so far.
The correctness of string-net condensation in our vacuum depend on resolving
the above problems.  Nature has four fascinating and somewhat strange
properties: gauge bosons, Fermi statistics, chiral fermions, and gravity.  The
string-net condensation picture provides a natural explanation for the first
two properties. Two more to go.

On the other hand, if we concern about a condensed matter problem: How to use
bosons to make artificial light and artificial fermions, then the string-net
picture and quantum order do provide an answer. To make artificial light and
artificial fermions, we simply let certain string-nets to condense.

In some recent work, quantum orders and their connection to emerging gauge
bosons and fermions were studied using PSG's, without realizing their
connection to string-net condensation.\cite{Wqoslpub,Wqoexct,Wlight} In this
paper, we will show that the quantum ordered states described by PSG's are
actually string-net condensed states. The gauge bosons and fermions produced
and protected by the PSG's have a very natural string-net
interpretation.\cite{Walight,LW0360} Quantum order, PSG, and string-net
condensation are different parts of the same story.  Here we will summarize
and expand those previous work and try to present a coherent picture for
quantum order, PSG, and string-net condensation, as well as the associated
emerging gauge bosons and fermions.

\subsection{Organization}

Section \ref{sec:Z2} reviews the work in \Ref{LW0360}. We will study an exactly
soluble spin-1/2 model on square lattice\cite{K9721,Wqoexct}.  The model was
solved using slave-boson approach.\cite{Wqoexct} This allowed us to identify
the PSG that characterizes the non-trivial quantum order in the ground
state.\cite{Wqoexct} Here, following \Ref{LW0360}, we will solve the model from
string-net condensation point of view.  Since the ground state of the model can be
described by both string-net condensation and PSG, this allows us to demonstrate
the direct connection between string-net condensation and PSG in section
\ref{PSGstring}.  The model is also one of the simplest models that
demonstrates the connection between string-net condensation and emerging gauge
field and fermions.\cite{K9721,LW0360}  

However, the above exact soluble model does not contain
gapless gauge boson and gapless fermions. If we regard the lattice scale as
the Planck scale, then gauge bosons and fermions do not ``exist'' in our model
in the sense discussed in section \ref{fq}.  In section \ref{masslessferm}, we
will discuss an exact soluble local bosonic model that contain massless Dirac
fermions.  In sections \ref{QED} and \ref{secQCD}, we will discuss local
bosonic models that give rise to massless electrons, quarks, gluons, and
photons.  Gauge bosons and fermions ``exist'' in those latter models.

\section{Local bosonic models}
\label{LBM}

In this paper, we will only consider local bosonic models.  Local bosonic
models are important since they are really local.  We note that a fermionic
model are in general non local since the fermion operators at different sites
do not commute, even when the sites are well separated. Due to their intrinsic
locality, local bosonic models are natural candidates for the fundamental
theory of nature.  In the following we will give a detailed definition of
local bosonic models.

To define a physical system, we need to specify (A) a total Hilbert space, (B)
a definition of a set of local physical operators, and (C) a Hamiltonian. With
this understanding, a local bosonic model is defined to be a model that
satisfies: (A) The total Hilbert space is a direct product of local Hilbert
spaces of finite dimensions. (B) Local physical operators are local bosonic
operators.  By definition, \emph{local bosonic operators} are operators acting
within a local Hilbert space or finite products of those operators for nearby
local Hilbert spaces.  Those operators are called local bosonic operators
since they all commute with each other when far apart. (C) The Hamiltonian is
a sum of local physical operators.

A spin-1/2 system on a lattice is an example of local bosonic models.  The
local Hilbert space is two dimensional which contains $|\up\>$ and $|\down\>$
states. Local physical operators are $\si^a_{\v i}$, $\si^a_{\v i}\si^b_{\v
i+\v x}$, \etc, where $\si^a$, $a=x,y,z$ are the Pauli matrices.

A free spinless fermion system (in 2 or higher dimensions) is not a
local bosonic model despite it has the same total Hilbert space as the
spin-1/2 system. This is because the fermion operators $c_{\v i}$ on different
sites do not commute and are not local bosonic operators. More importantly,
the fermion hoping Hamiltonian in 2 and higher dimensions cannot be written as
a sum of local bosonic operators. (Note in higher dimensions, we cannot write
all the hoping terms $c_{\v i}^\dag c_{\v j}$ as product of local bosonic
operators. However, due to the Jordan-Wigner transformation, a 1D fermion
hoping $c^\dag_{i+1}c_i$ can be written as a local bosonic operators. Hence, a
1D fermion system can be a local bosonic model if we exclude $c_{\v i}$ from
our definition of local physical operators.) 

The bosonic field theory without cut-off is not a local bosonic model. This is
because the local Hilbert space does not have a finite dimension.
A lattice gauge theory is not a  local bosonic model. This is because its
total Hilbert space cannot be a direct product of local Hilbert spaces.

Another counter example of local bosonic model is a quantum closed-string-net
model. A quantum closed-string-net model on lattice can be defined in the
following way.  Let us consider only strings that cover nearest neighbor
links.  A closed-string configuration may have many closed strings with or
without overlap. We will call a closed-string configuration a closed
string-net.  For every closed string-net , we assign a quantum state.  All
those quantum states form a basis of the total Hilbert space of the 
closed-string-net model. Just like lattice gauge theory, the 
closed-string-net model is not a local bosonic model since the total Hilbert
space cannot be a direct product of local Hilbert spaces. It turns out that
closed-string-net models and lattice gauge model are closely related.
In fact some closed-string-net models (or statistical membrane models)
are equivalent to lattice gauge models.\cite{KS7595,BMK7793,S8053,ID89}

\section{$Z_2$ spin liquid and string-net condensation on square lattice}
\label{sec:Z2}

\subsection{Hamiltonians with closed-string-net condensation}

Let us first consider an arbitrary spin-1/2 model on a square lattice.  The
first question that we want to ask is what kind of spin interaction can given
rise to a low energy gauge theory. If we believe the connection between gauge
theory and closed-string-net theory,\cite{KS7595,BMK7793,S8053,ID89,Walight}
then one way to obtain a low energy gauge theory is to design a spin
interaction that allow strong fluctuations of large closed string-nets, but
forbid other types of fluctuations (such as local spin flip, open string-net
fluctuations, \etc).  (Note that closed string-nets are nets of strings formed
by intersecting/overlapping closed strings, while open string-nets are nets of
strings containing at least an open string.) We hope the presence of strong
fluctuations of large closed-strings will lead to condensation of closed
strings of arbitrary sizes, which in turn gives rise to a low energy gauge
theory.

\begin{figure}
\centerline{
\includegraphics[width=2.5in]{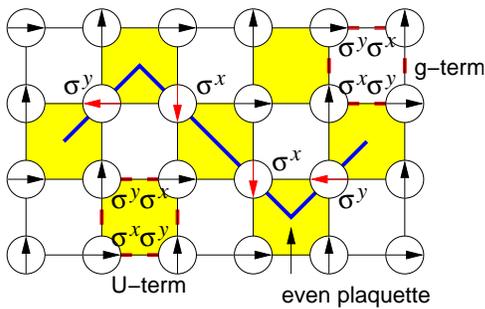}
}
\caption{
An open-string excitation on top of the ground state of $H_J$.
}
\label{string}
\end{figure}

Let us start with
\begin{equation}
 H_J = - J\sum_{even} \si^{x}_{\v i} - J\sum_{odd} \si^{y}_{\v i}
\end{equation}
where $\v i=(i_x,i_y)$ labels the lattice sites, $\si^{x,y,z}$ are the Pauli
matrices, and $\sum_{even}$ (or $\sum_{odd}$) is a sum over even sites with
$(-)^{\v i}\equiv (-1)^{i_x+i_y}=1$ (or over odd sites with $(-)^{\v i}\equiv
(-1)^{i_x+i_y}=-1$). The ground state of $H_J$, $|0\>$, has spins pointing to
$x$-direction on even sites and to $y$-direction on odd sites (see Fig.
\ref{string}). Such a state will be defined as a state with no string.

To create a string excitation, we first draw a string that connect nearest
neighbor \emph{even} plaquettes (see Fig. \ref{string}). We then flip the
spins in the string.  Such a string state is created by the following string
creation operator (or simply, string operator):
\begin{equation}
\label{stringOP}
 W(C) = \prod_C \si^{a_{\v i}}_{\v i}
\end{equation}
where the product $\prod_C$ is over all the sites on the string, $a_{\v i}=y$
if $\v i$ is even and $a_{\v i}=x$ if $\v i$ is odd.  A generic string state
has a form
\begin{equation}
 |C_1C_2 ...\> = W(C_1)W(C_2)...|0\>
\end{equation}
where $C_1$, $C_2$, ... are strings with no overlapping ends.  
Such a state will be called a string-net state and 
\begin{equation*}
W(C_{net})=W(C_1)W(C_2)... 
\end{equation*}
will be called a string-net operator.  The state $|C_1C_2 ...\>$ is an
open-string-net state if at least one of $C_i$ is an open string. The
corresponding operator $W(C_{net})$ will be called an open-string-net operator.
If all $C_i$ are closed loops, then $|C_1C_2 ...\>$ is an closed-string-net
state and $W(C_{net})$ an closed-string-net operator.  The Hamiltonian has no
string-net condensation since its ground state $|0\>$ contains no string-nets.
To obtain a Hamiltonian with closed-string-net condensation, we need to first
find a Hamiltonian whose ground state contains a lot of closed string-nets of
arbitrary sizes and do not contain open string-nets.

Let us first write down a Hamiltonian such that closed strings cost no energy
and any open strings cost a large energy.  One such Hamiltonian has a form
\begin{align}
 H_U =& -U \sum_{even} \hat F_{\v i}  \nonumber\\
\hat F_{\v i} =& 
\si^x_{\v i}
\si^y_{\v i+\v x}
\si^x_{\v i+\v x+\v y}
\si^y_{\v i+\v y}
\end{align}
We find the no-string state $|0\>$ is one of the ground state of $H_U$
(assuming $U>0$) with energy $-UN_{site}$. All the closed-string-net states, 
such
as $W(C_{close})|0\>$, are also ground state of $H_U$ since
$[H_U,W(C_{close})]=0$. 
An open-string state $W(C_{open})|0\>$ is also an eigenstate of $H_U$ but with
energy $-UN_{site}+2U$. We see that each end of open string cost an energy
$U$. We also note that the energy of closed strings does not depend on the
length of closed strings. Thus the closed strings in $H_U$ have no tension. We
can introduce a string tension by adding the $H_J$ to our Hamiltonian. The
string tension will be $2J$ per site (or per segment).  We note that, any
string-net state $|C_1C_2 ...\>$ is an eigenstate of $H_U+H_J$.  Thus,
string-nets in the model
described by $H_U+H_J$ do not fluctuate and hence cannot condense.  To make
string-nets to fluctuate, we need a $g$-term
\begin{equation}
 H_g = g \sum_{\v p} U(C_{\v p})
\end{equation}
where $\v p$ labels the odd plaquettes and $C_{\v p}$ is the closed string
around the plaquette $\v p$.  In fact
\begin{equation}
 H_g = -g \sum_{odd} \hat F_{\v i}
\end{equation}
This way, we obtain the Hamiltonian of our spin-1/2 model
\begin{equation}
\label{HUJg}
 H = H_U+H_J+H_g
\end{equation}

\subsection{String condensation and low energy effective theory}

When $J=0$ in \Eq{HUJg}, 
the model is exactly soluble since $[\hat F_{\v i},\hat F_{\v
j}]=0$.\cite{K9721,Wqoexct} All the eigenstates of $H_U+H_g$ can be obtained
from the common eigenstates of $\hat F_{\v i}$. Since $\hat F_{\v i}^2=1$, the
eigenvalues of $\hat F_{\v i}$ are simply $\pm 1$. Thu
s all the eigenstates of
$H_U+H_g$ are labeled by $\pm 1$ on each plaquette. (Note, this is not true
for finite systems where the boundary condition introduce additional
complications.\cite{Wqoexct}) The energies of those eigenstates are sum of
eigenvalues of $\hat F_{\v i}$ weighted by $U$ and $g$.

\begin{figure}
\centerline{
\includegraphics[width=2in]{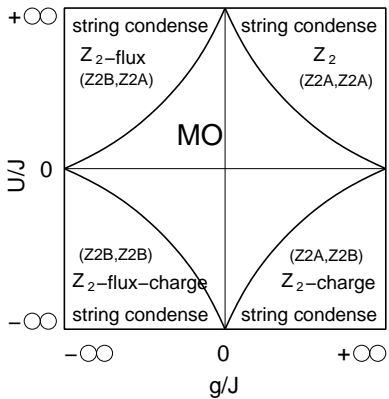}
}
\caption{
The proposed phase diagram for the $H=H_U+H_g+H_J$ model. $J$ is assumed to be
positive. The four string-net condensed phases are characterized by a pair of
PSG's $(PSG_{charge},PSG_{vortex})$.
MO marks an magnetic ordered state.
}
\label{phase}
\end{figure}

From the results of exact soluble model, we suggest a phase diagram of our
model as sketched in Fig. \ref{phase}.  We will show that the phase diagram
contains four different string-net condensed phases and one phase with no
string condensation.  All the phases have the same symmetry and are
distinguished only by their different quantum orders.

Let us first discuss the phase with $U,g>0$. We will assume $J=0$ and $U\gg
g$. In this limit, all states containing open strings will have an energy of
order $U$. The low energy states contain only closed strings (or more generally
closed string-nets) and satisfy
\begin{equation}
 \hat F_{\v i}|_{\v i=even} = 1
\end{equation}
For infinite systems, the different low energy states are labeled by
the eigenvalues of $\hat F_{\v i}$ on odd plaquettes:
\begin{equation}
 \hat F_{\v i}|_{\v i=odd} = \pm 1
\end{equation}
In particular, the ground state is given by
\begin{equation}
\hat F_{\v i}|_{\v i=odd} = 1. 
\end{equation}

All the closed-string-net operators $W(C_{net})$ commute with $H_U+H_g$. Hence
the ground state $|\Psi_0\>$ of $H_U+H_g$ satisfies 
\begin{equation}
\<\Psi_0| W(C_{net})|\Psi_0\> =1.  
\end{equation}
Thus the $U,g>0$ ground state has a closed-stringi-net condensation.
The low energy excitations above the ground state can be obtained by flipping
$\hat F_{\v i}$ from $1$ to $-1$ on some odd plaquettes.

If we view  $\hat F_{\v i}$ on odd plaquettes as the flux in $Z_2$ gauge
theory, we find that the low energy sector of model is identical to a $Z_2$
lattice gauge theory, at least for infinite systems. This suggests that the
low energy effective theory of our model is a $Z_2$ lattice gauge theory.

However, one may object this result by pointing out that the low energy sector
of our model is also identical to an Ising model with one spin on each the odd
plaquette. Thus the the low energy effective theory should be the Ising model.
We would like to point out that although the low energy sector of our model is
identical to an Ising model for infinite systems, the low energy sector of our
model is different from an Ising model for finite systems.  For example, on a
finite even by even lattice with periodic boundary condition, the ground state
of our model has a four-fold degeneracy.\cite{K9721,Wqoexct} The Ising model
does not have such a degeneracy.  Also, our model contains an excitation that
can be identified as $Z_2$ charge (see below). Therefore, the low energy
effective theory of our model is a $Z_2$ lattice gauge theory instead of an
Ising model.  The $\hat F_{\v i}=-1$ excitations on odd plaquettes can be
viewed as the $Z_2$ vortex excitations in the $Z_2$ lattice gauge theory.

\begin{figure}
\centerline{
\includegraphics[width=2.2in]{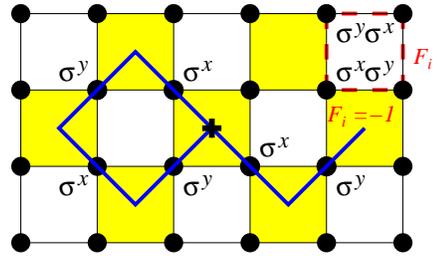}
}
\caption{
A hopping of the $Z_2$ charge around four nearest neighbor even plaquettes.
}
\label{hop}
\end{figure}

\subsection{Three types of strings and emerging fermions}
\label{3type}

What is the $Z_2$ charge excitations?  We note that, in the closed-string-net
condensed state, the action of the closed-string operator \Eq{stringOP} on the
ground state is trivial.  This suggests that the action of the open-string
operators on the ground state only depend on the ends of strings, since two
open strings with the same ends only differ by a closed string.  Therefore, an
open-string operator create two particles at its ends when acting on the
string condensed state.  Since the strings in \Eq{stringOP} only connect even
plaquettes, the particle corresponding to the ends of the open strings always
live on the even plaquettes. We will call such a string T1 string.  Form the
commutation relation between $\hat F_{\v i}$ and the open-string operators, we
find that the open-string operators flip the sign of $\hat F_{\v i}$ at its
ends.  Thus each particle created by the open-string operators has an energy
$2U$.  Now, let us consider the hopping of one such particle around four
nearest neighbor even plaquettes (see Fig.  \ref{hop}). We see that the
product of the the four hopping amplitudes is given by the eigenvalue of $\hat
F_{\v i}$ on the odd plaquette in the middle of the four even
plaquettes.\cite{K9721,LW0360} This is exactly the relation between charge and
flux.  Thus if we identify $\hat F_{\v i}$ on odd plaquettes as $Z_2$ flux,
then the ends of strings on even plaquettes will correspond to the $Z_2$
charges.  We note that, due to the closed-string condensation, the ends of
open strings are not confined and have only short ranged interactions between
them.  Thus the $Z_2$ charges behave like quasiparticles with no string
attached.

Just like the $Z_2$ charges, a pair $Z_2$ vortices is also created by an open
string operator.  Since the $Z_2$ vortices correspond to flipped $\hat F_{\v
i}$ on \emph{odd} plaquettes, the open-string operator that create $Z_2$
vortices is also given by \Eq{stringOP}, except now the product is over a
string that connect \emph{odd} plaquettes.  We will call such a string T2
string. (The strings connecting \emph{even} plaquettes were called T1
strings.)

We would like to point out that the reference state (\ie the no string state)
for the T2 string is different from that of the T1 string.  The no-T2-string
state is given by $|\t 0\>$ with spin pointing in $y$-direction on even sites
and $x$-direction on odd sites. Since the T1 and T2 strings have different
reference state, we cannot have a dilute gas of the T1 strings and the T2
strings at the same time.  One can easily check that the T2 string operators
also commute with $H_U+H_g$. Therefore, the ground state $|\Psi_0\>$, in
addition to the T1 closed-string condensation, also has a T2 closed-string
condensation.  

The hopping of a $Z_2$ vortex is induced by a short T2 open-string.  Since the
T2 open-strings operators all commute with each other, the $Z_2$ vortex behave
like bosons.  Similarly, the $Z_2$ charges also behave like bosons. However,
T1 open-string operators and T2 open-string operators do not commute.  As a
result, the ends of T1 string and the ends of T2 string have nontrivial mutual
statistics.  As we have already shown that moving a $Z_2$ charge around a
$Z_2$ vortex generate a phase $\pi$, the $Z_2$ charges  and the $Z_2$ vortices
have a semionic mutual statistics.

The T3 strings are defined as bound states of T1 and T2 strings.  The T3
string operator has a form $ W(C) = \prod_{n} \si^{l_n}_{\v i_n}$.  where $C$
is a string connecting the mid-points of the neighboring links (see Fig.
\ref{fhop}), and $\v i_n$ are sites on the string.  $l_m=z$ if the string does
not turn at site $\v i_m$ (see Fig. \ref{fhop}).  $l_m=x$ or $y$ if the string
makes a turn at site $\v i_m$.  $l_m=x$ if the turn forms a upper-right or
lower-left corner.  $l_m=y$ if the turn forms a lower-right or upper-left
corner.  (See Fig. \ref{fhop}.) The ground state also has a condensation of T3
closed-strings. The ends of T3 string, as bound states of the $Z_2$ charges
and the $Z_2$ vortices, are \emph{fermions}.  The bound state is formed by a
$Z_2$ charge and a $Z_2$ vortex on the two plaquettes on the two side of a
link (\ie $F_{\v i}=-1$ on the two sides of the link).  Thus the fermions live
on the links. It is interesting to see that string-net condensation in our
model directly leads to $Z_2$ gauge structure and three new type of
quasiparticles: $Z_2$ charge, $Z_2$ vortex, and fermions. Fermions, as ends of
open T3 strings, emerge from our purely bosonic model.

Since ends of T1 string are $Z_2$ charges, the T1 string can be viewed as
strings of $Z_2$ ``electric'' flux. Similarly, the T2 string can be viewed as
strings of $Z_2$ ``magnetic'' flux.

\section{Classification of different string condensations by PSG}
\label{PSGstring}

\subsection{Four classes of string-net condensations}

As we have seen in last section that when $U>0$, $g>0$, and $J=0$, the ground
state of our model is given by
\begin{align}
\hat F_{\v i}|_{\v i=even} =& 1,  &
\hat F_{\v i}|_{\v i=odd} =& 1. 
\end{align}
We will call such a phase $Z_2$ phase to stress the low energy $Z_2$ gauge
structure. In the $Z_2$ phase, the T1 string operator $W_1(C_1)$ and the T2
string operator $W_2(C_2)$ have the following expectation values
\begin{equation}
 \< W_1(C_1)\>=1,\ \ \ \< W_2(C_2)\>=1
\end{equation}

When $U>0$, $g<0$, and $J=0$, the ground state is given by
\begin{align}
\hat F_{\v i}|_{\v i=even} =& 1,  &
\hat F_{\v i}|_{\v i=odd} =& -1. 
\end{align}
We see that there is $\pi$-flux through each odd plaquette. We will call such
a phase $Z_2$-flux phase. The T1 string operator and the T2 string operator
have the following expectation values
\begin{equation}
 \< W_1(C_1)\>=(-)^{N_{odd}},\ \ \ \< W_2(C_2)\>=1
\end{equation}
where $N_{odd}$ is the number of odd-plaquettes enclosed by the T1 string
$C_1$.

When $U<0$, $g>0$, and $J=0$, the ground state is 
\begin{align}
\hat F_{\v i}|_{\v i=even} =& -1,  &
\hat F_{\v i}|_{\v i=odd} =& 1. 
\end{align}
The ground state has a $Z_2$ charge on each even plaquette. We will call such
a phase $Z_2$-charge phase. The T1 string operator and the T2 string operator
have the following expectation values
\begin{equation}
 \< W_1(C_1)\>=1,\ \ \ \< W_2(C_2)\>=(-)^{N_{even}}
\end{equation}
where $N_{even}$ is the number of even-plaquettes enclosed by the T2 string
$C_2$. Note that the $Z_2$-flux phase and the $Z_2$-charge phase, different
only by a lattice translation, are essentially the same phase.

When $U<0$, $g<0$, and $J=0$, the ground state becomes
\begin{align}
\hat F_{\v i}|_{\v i=even} =& -1,  &
\hat F_{\v i}|_{\v i=odd} =& -1. 
\end{align}
There is a $Z_2$ charge on each even plaquette and $\pi$-flux through each odd
plaquette. We will call such a phase $Z_2$-flux-charge phase.  The T1 string
operator and the T2 string operator have the following expectation values
\begin{equation}
 \< W_1(C_1)\>=(-)^{N_{odd}},\ \ \ \< W_2(C_2)\>=(-)^{N_{even}}
\end{equation}

\subsection{PSG and ends of condensed strings}
\label{PSGend}

From the different $\< W_1(C_1)\>$ and $\< W_2(C_2)\>$, we see that the above
four phases have different string-net condensations. However, they all have
the same symmetry.  This raises an issue. Without symmetry breaking, how do we
know the above four phases are really different phases? How do we know that it
is impossible to change one string-net condensed state to another without a
phase transition?

In the following, we will show that the different string-net condensations can
be described by different PSG's (just like different symmetry breaking orders
can be described by different symmetry groups of ground states.) In
\Ref{Wqoslpub,Wqogen}, different quantum orders were introduced via their
different PSG's. The connection between string-net condensation and PSG allows
us to connect string-net condensation to the quantum order introduced in
\Ref{Wqoslpub,Wqogen}.  In particular, the PSG's are shown to be a universal
property of a quantum phase, which can be changed only by phase transitions.
Thus the different PSG's for the different string-net condensed states
indicate that those different string-net condensed states belong to different
quantum phases.

When closed-string-nets condense, the ends of open strings behave like
independent particles. Let us consider two particles states $|\v p_1\v p_2\>$
described by the two ends of a T1 string. Note that the ends of the T1
strings, and hence the $Z_2$ charges, only live on the even plaquettes. Here
$\v p_1$ and $\v p_2$ label the even plaquettes. For our model $H_U+H_g$, $|\v
p_1\v p_2\>$ is an energy eigenstate and the $Z_2$ charges do not hop. Here we
would like to add a term 
\begin{equation}
H_t=t\sum_{\v i} (\si^x_{\v i} +\si^y_{\v i}) +t'\sum_{\v i} \si^z_{\v i}
\end{equation}
to the Hamiltonian. The $t$-term $t\sum_{\v i} (\si^x_{\v i} +\si^y_{\v i})$
makes the $Z_2$ charges to hop among the even plaquettes with a hopping
amplitude of order $t$. The dynamics of the two $Z_2$ charges is described by
the following effective Hamiltonian in the two-particle Hilbert space
\begin{equation}
 H = H(\v p_1) + H(\v p_2)
\end{equation}
where $H(\v p_1)$ describes the hopping of the first particle $\v p_1$ and
$H(\v p_2)$ describes the hopping of the second particle $\v p_2$.  Now we can
define the PSG in a string-net condensed state. The PSG is nothing but the
symmetry group of the hopping Hamiltonian $H(\v p)$.

Due to the translation symmetry of the underlying model $H_U+H_g+H_t$, we
may naively expect the hopping Hamiltonian of the $Z_2$ charge $H(\v p)$ also 
have a translation symmetry
\begin{align}
\label{HTxy}
 H(\v p)=& T_{xy}^\dag H(\v p)T_{xy},\ \ \
 T_{xy}|\v p\> = |\v p+\v x+\v y\> \nonumber\\
 H(\v p)=& T_{x\bar y}^\dag H(\v p)T_{x\bar y},\ \ \
 T_{x\bar y}|\v p\> = |\v p+\v x-\v y\> 
\end{align}
The above implies PSG = translation symmetry group.  It turns out that
\Eq{HTxy} is too strong. The underlying spin model can have translation
symmetry even when $H(\v p)$ does not satisfy \Eq{HTxy}.  However, the
possible symmetry groups of $H(\v p)$ (the PSG's) are strongly constrained by
the translation symmetry of the underlying spin model. In the follow, we will
explain why the PSG can be different from the symmetry group of the physical
spin model, and what conditions that the PSG must satisfy in order to be
consistent with the translation symmetry of the spin model.

We note that a string
always has two ends. Thus a physical state always has an even number of $Z_2$
charges. The actions of translation on a two-particle state are given by
\begin{align}
 T^{(2)}_{xy}|\v p_1,\v p_2\> =& e^{\th_{xy}(\v p_1,\v p_2)}
|\v p_1+\v x+\v y, \v p_2+\v x+\v y\>
\nonumber\\
 T^{(2)}_{x\bar y}|\v p_1,\v p_2\> =&e^{\th_{x\bar y}(\v p_1,\v p_2)} 
|\v p_1+\v x-\v y, \v p_2+\v x-\v y\>
\end{align}
The phases $e^{\th_{xy}(\v p_1,\v p_2)}$ and $e^{\th_{x\bar y}(\v p_1,\v
p_2)}$ come from the ambiguity of the location of the string that connect $\v
p_1$ and $\v p_2$.  \ie the  phases can be different if the string connecting
the two $Z_2$ charges has different locations.  $T^{(2)}_{xy}$ and
$T^{(2)}_{x\bar y}$ satisfy the algebra of translations
\begin{equation}
\label{T2alg}
 T^{(2)}_{xy}T^{(2)}_{x\bar y} = T^{(2)}_{x\bar y} T^{(2)}_{xy}
\end{equation}
$T^{(2)}_{xy}$ and $T^{(2)}_{x\bar y}$ are direct products of translation
operators on the single-particle states. Thus, in some sense, the
single-particle translations are square roots of two-particle translations.

The most general form of single-particle translations is given by
$T_{xy}G_{xy}$ and $T_{x\bar y}G_{x\bar y}$, where the actions of operators
$T_{xy,x\bar y}$ and $G_{xy,x\bar y}$ are defined as
\begin{align}
 T_{xy}|\v p\> =& |\v p+\v x+\v y\> \nonumber\\
 T_{x\bar y}|\v p\> =& |\v p+\v x-\v y\> \nonumber\\
 G_{xy}|\v p\> =& e^{i\phi_{xy}(\v p)} |\v p\> \nonumber\\
 G_{x\bar y}|\v p\> =& e^{i\phi_{x\bar y}(\v p)} |\v p\>
\end{align}
In order for the direct product $T^{(2)}_{xy} = T_{xy}G_{xy} \otimes
T_{xy}G_{xy}$ and $T^{(2)}_{x\bar y} = T_{x\bar y}G_{x\bar y} \otimes T_{x\bar
y}G_{x\bar y}$ to reproduce the translation algebra \Eq{T2alg}, we only
require $T_{xy}G_{xy}$ and $T_{x\bar y}G_{x\bar y}$ to satisfy
\begin{equation}
\label{Talg1}
T_{xy}G_{xy}T_{x\bar y}G_{x\bar y} =
T_{x\bar y}G_{x\bar y} T_{xy}G_{xy}
\end{equation}
or
\begin{equation}
\label{Talg2}
T_{xy}G_{xy}T_{x\bar y}G_{x\bar y} = -
T_{x\bar y}G_{x\bar y} T_{xy}G_{xy}
\end{equation}
The operators $T_{xy}G_{xy}$ and $T_{x\bar y}G_{x\bar y}$ generate a group.
Such a group is the PSG introduced in \Ref{Wqoslpub}. The two different
algebra \Eq{Talg1} and \Eq{Talg2} generate two different PSG's, both are
consistent with the translation group acting on the two-particle states. We
will call the PSG generated by \Eq{Talg1} $Z2A$ PSG and the PSG generated by
\Eq{Talg2} $Z2B$ PSG.

Let us give a more general definition of PSG. A PSG is a group. It is a
extension of symmetry group (SG), \ie a PSG contain a normal subgroup (called
invariant gauge group or IGG) such that
\begin{equation}
 PSG/IGG = SG
\end{equation}

For our case, the SG is the translation group $SG=\{1,T^{(2)}_{xy},
T^{(2)}_{x\bar y},...\}$.
For every element in SG, $a^{(2)} \in SG$, there are one or several elements
in PSG, $a\in PSG$, such that $a\otimes a = a^{(2)}$.  The IGG in our PSG is
formed by the transformations $G_0$ on the singe-particle states that satisfy
$G_0\otimes G_0=1$. We find that IGG is generated by
\begin{equation} 
G_0|\v p\>=-|\v p>
\end{equation} 
$G_0$, $T_{xy}G_{xy}$ and $T_{x\bar y}G_{x\bar y}$ generate the $Z2A$ and
$Z2B$ PSG's.

Now we see that the underlying translation symmetry does not require the
single-particle hopping Hamiltonian $H(\v p)$ to have a translation symmetry.
It only require $H(\v p)$ to be invariant under the $Z2A$ PSG or the $Z2B$
PSG. When $H(\v p)$ is invariant under the $Z2A$ PSG, the hopping Hamiltonian
has the usual translation symmetry. When $H(\v p)$ is invariant under the
$Z2B$ PSG, the hopping Hamiltonian has a magnetic translation symmetry
describing a hopping in a magnetic field with $\pi$-flux through each odd
plaquette.

\subsection{PSG's classify different string-net condensations}

After understand the possible PSG's for the hopping Hamiltonian of the ends
of strings, now we are ready to calculate the actual PSG's.  Let us
consider two ground states of our model $H_U+H_g+H_t$. One has $\hat F_{\v
i}|_{\v i=odd}=1$ (for $g>0$) and the other has $\hat F_{\v i}|_{\v i=odd}=-1$
(for $g<0$).  Both ground states have the same translation symmetry in $\v
x+\v y$ and $\v x-\v y$ directions. However, the corresponding single-particle
hopping Hamiltonian $H(\v p)$ has different symmetries.  For the $\hat F_{\v
i}|_{\v i=odd}=1$ state, there is no flux through odd plaquettes and $H(\v p)$
has the usual translation symmetry. It is invariant under the $Z2A$ PSG. While
for the $\hat F_{\v i}|_{\v i=odd}=-1$ state, there is $\pi$-flux through odd
plaquettes and $H(\v p)$ has a magnetic  translation symmetry. Its PSG is the
$Z2B$ PSG.  Thus the $\hat F_{\v i}|_{\v i=odd}=1$ state and the $\hat F_{\v
i}|_{\v i=odd}=-1$ state have different orders despite they have the same
symmetry.  The different quantum orders in the two states can be characterized
by their different PSG's.

\begin{figure}
\centerline{
\includegraphics[width=2in]{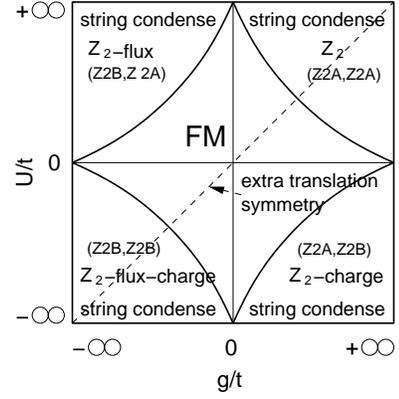}
}
\caption{
The proposed phase diagram for the $H=H_U+H_g+H_t$ model. 
$t=t'$ is assumed to be
positive. The four string-net condensed phases are characterized by a pair of
PSG's $(PSG_{charge},PSG_{vortex})$. FM marks a ferromagnetic phase.
}
\label{phaseUgt}
\end{figure}

The above discussion also apply to the $Z_2$ vortex and T2 strings. Thus the
quantum orders in our model are described by a pair of PSG's $(PSG_{charge},
PSG_{vortex})$, one for the $Z_2$ charge and one for the $Z_2$ vortex.  The
PSG pairs $(PSG_{charge}, PSG_{vortex})$ allows us to distinguish four
different string-net condensed states of model $H=H_U+H_g+H_t$.(See Fig.
\ref{phaseUgt}.) 

Now let us assume $U=g$ in our model:
\begin{equation}
\label{HV}
H_U+H_g+H_t = H_t -V \sum_{\v i} \hat F_{\v i} 
\end{equation}
The new physical spin model has a larger translation symmetry generated by
$\Del \v i=\v x$ and $\Del \v i=\v y$ (see Fig. \ref{phaseUgt}). Due to the
enlarged symmetry group, the quantum orders in the new system should be
characterized by a new PSG. In the following, we will calculate the new PSG.

The single-particle states are given by $|\v p\>$. When $\v p$ is even, $|\v
p\>$ corresponds to a $Z_2$ charge and when $\v p$ is odd, $|\v p\>$
corresponds to a $Z_2$ vortex. We see that a translation by $\v x$ (or $\v y$)
will change a $Z_2$ charge to a $Z_2$ vortex or a $Z_2$ vortex to a $Z_2$
charge. Therefore the effective  single-particle hopping Hamiltonian $H(\v p)$
only contain hops between even plaquettes or odd plaquettes.  The
single-particle Hamiltonian $H(\v p)$ is invariant under the following two
transformations $G_0$ and $G_0'$:
\begin{equation}
G_0|\v p\>=-|\v p>, \ \ \
G_0'|\v p\>=(-)^{\v p}|\v p>
\end{equation}
We note that $G_0\otimes G_0= G_0'\otimes G_0'=1$. Therefore both $G_0$ and
$G_0'$ correspond to the identity element of the symmetry group of
two-particle states.  $(G_0,G_0')$ generate the IGG of the new PSG.
The new IGG is $Z_2\times Z_2$.

The translations of single-particle states by $\v x$ and by $\v y$
are generated by $T_xG_x$ and $T_yG_y$. The translation
by $\v x+\v y$ and by $\v x-\v y$ are given by
\begin{align}
 T_{xy}G_{xy} =& T_yG_y T_xG_x \nonumber\\
 T_{x\bar y}G_{x\bar y} =& (T_yG_y)^{-1} T_xG_x 
\end{align}
Since $T_{xy}G_{xy}$ and $T_{x\bar y}G_{x\bar y}$ are the translations of the
$Z_2$ charge and the $Z_2$ vortex discussed above, we find 
\begin{equation}
\label{TxyTxyeta}
(T_{x\bar y}G_{x\bar y})^{-1} (T_{xy}G_{xy})^{-1}
T_{x\bar y}G_{x\bar y} T_{xy}G_{xy}  =\eta
\end{equation}
where $\eta =1$ for the $(Z2A, Z2A)$ state with $\hat F_{\v i}=1$
and $\eta =-1$ for the $(Z2B, Z2B)$ state with $\hat F_{\v i}=-1$.
Also $T_xG_x$ and $T_yG_y$ must satisfy
\begin{equation}
\label{TxTyIGG}
(T_yG_y)^{-1} (T_xG_x)^{-1} T_yG_y T_xG_x \in IGG
\end{equation}
since on the two-particle states
\begin{equation}
(T^{(2)}_y)^{-1} (T^{(2)}_x)^{-1} T^{(2)}_y T^{(2)}_x =1
\end{equation}
Therefore, $(T_yG_y)^{-1} (T_xG_x)^{-1} T_yG_y T_xG_x$ may take the following
possible values $1$, $-1$, $(-)^{\v p}$, and $-(-)^{\v p}$.
Only choices $\eta^{\v p}$ and $-\eta^{\v p}$ are consistent with
\Eq{TxyTxyeta} and we have
\begin{equation}
\label{TxTyeta}
(T_yG_y)^{-1} (T_xG_x)^{-1} T_yG_y T_xG_x = \eta'\eta^{\v p}
\end{equation}
We like to point out that the different choices of $\eta'=\pm 1$ do not lead
to different PSG's. This is because if $T_xG_x$ is a symmetry of the $H(\v
p)$, then $T_xG_x(-)^{\v p}$ is also a symmetry of the $H(\v
p)$. However, the change $G_x\to G_x(-)^{\v p}$ will change the sign of
$\eta'$. Thus $\eta'=1$ and $\eta'=-1$ will lead to the same PSG. But
the different signs of $\eta$ will lead to different PSG's.

$(G_0,G_0')$ and $(T_xG_x, T_yG_y)$ generate the new PSG.  The single-particle
Hamiltonian $H(\v p)$ is invariant under such a PSG.  $\eta=1$ and $\eta=-1$
correspond to two different PSG's that characterize two different quantum
orders. The ground state for $V>0$ and $|V|\gg t$ (see \Eq{HV}) is described
by the $\eta=1$ PSG.  The ground state for $V<0$ and $|V|\gg t$ is described
by the $\eta=-1$ PSG. The two ground states have different quantum orders and
different string-net condensations.

\subsection{Different PSG's from the ends of different condensed strings}

In this section we still assume $U=g$ and consider only the translation
invariant model \Eq{HV}.
In the above we discussed the PSG for the ends of one type of condensed
strings in different states.  In this section, we will concentrate on only one
ground state. We know that the ground state of our spin-1/2 model contain
condensations of several type of strings. We like to calculate the the
different PSG's for the different condensed strings.

The PSG's for the condensed T1 and T2 strings were obtained above.  Here we
will discuss the PSG for the T3 string.  Since the ends of the T3 strings live
on the links, the corresponding single-particle hopping Hamiltonian $H_f(\v
l)$ describes fermion hopping between links.  Clearly, the symmetry group (the
PSG) of $H_f(\v l)$ can be different from that of $H(\v p)$.

\begin{figure}
\centerline{
\includegraphics[width=2.2in]{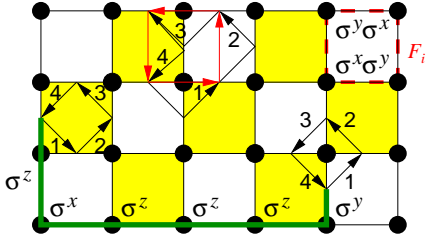}
}
\caption{
Fermion hopping around a plaquette, around a square, and around a
site.
}
\label{fhop}
\end{figure}

Let us consider fermion hopping around some small loops. The four hops of a
fermion around a site $\v i$ (see Fig. \ref{fhop}) are generated by $\si^y_{\v
i}$, $\si^x_{\v i}$, $\si^y_{\v i}$, and $\si^x_{\v i}$.  The total amplitude
of a fermion hopping around a site is $\si^y_{\v i}\si^x_{\v i}\si^y_{\v
i}\si^x_{\v i}=-1$. The fermion hopping around a site always sees $\pi$-flux.
The four hops of a fermion around a plaquette $\v p$ (see Fig. \ref{fhop}) are
generated by $\si^x_{\v i_0}$, $\si^y_{\v i_0+\v x}$, $\si^x_{\v i_0+\v x+\v
y}$, and $\si^y_{\v i_0+\v y}$, where $\v i_0$ is the lower left corner of the
plaquette $\v p$. The total amplitude of a fermion hopping around a plaquette
is given by $\si^y_{\v i_0+\v y} \si^x_{\v i_0+\v x+\v y} \si^y_{\v i_0+\v x}
\si^x_{\v i_0} =\hat F_{\v i_0}$. When $V>0$, the ground state has $\hat F_{\v
i}=1$. However, since site $\v i_0$ is next to the end of T3 string, we have
$\hat F_{\v i_0}=-\hat F_{\v i}=-1$. In this case, the fermion hopping around a
plaquette sees $\pi$-flux. For $V<0$ ground state, we find that
fermion hopping around a plaquette sees no flux.

Let us define the fermion hopping $\v l\to \v l+\v x$ as the combination of
two hops
$\v l\to \v l+\frac{\v x}{2} -\frac{\v y}{2}\to \v l+\v x$
and the fermion hopping $\v l\to \v l+\v y$ as the combination of 
$\v l\to \v l+\frac{\v x}{2} +\frac{\v y}{2}\to \v l+\v y$ (see Fig.
\ref{fhop}). Under such a definition, a fermion hopping around a square $\v
l\to \v l+\v x\to \v l+\v x+\v y\to \v l +\v y \to \v l$ correspond to a
fermion hopping around a site and a fermion hopping around a plaquette
discussed above (see Fig. \ref{fhop}). Therefore, the total amplitude for a
fermion hopping around a square is given by the sign of $V$: $\sgn(V)$. 
We find the translation
symmetries $(T_xG_x, T_yG_y)$ of the fermion hopping $H_f(\v l)$ satisfies
\begin{equation}
\label{TxTyetaf}
(T_yG_y)^{-1} (T_xG_x)^{-1} T_yG_y T_xG_x = \sgn(V)
\end{equation}
which is different from the translation algebra for $H(\v p)$ \Eq{TxTyeta}.
$H_f(\v l)$ is also invariant under $G_0$:
\begin{equation}
G_0|\v l\> = -|\v l\> 
\end{equation}
$(G_0,T_xG_x,T_yG_y)$ generate the symmetry group - the fermion PSG - of
$H_f(\v l)$.  We will call the fermion PSG \Eq{TxTyetaf}
for $\sgn(V)=1$ the $Z2A$ PSG and the
fermion PSG for $\sgn(V)=-1$ the $Z2B$ PSG. We see that the quantum orders in
the ground state can also be characterized using the fermion PSG. The quantum
order in the $V>0$ ground state is characterized by the $Z2A$ PSG and the
quantum order in the $V<0$ ground state is characterized by the $Z2B$ PSG.

In \Ref{Wqoexct}, the spin-1/2 model \Eq{HV} (with $t=t'=0$) was viewed as a
hardcore boson model. The model was solved using slave-boson approach by
splitting a boson into two fermions. Then it was shown the fermion hopping
Hamiltonian for $V>0$ and $V<0$ states have different symmetries, or invariant
under different PSG's. According to the arguments in \Ref{Wqoslpub}, the
different PSG's imply different quantum orders in the $V>0$ and $V<0$ ground
state states.  The PSG's obtained in \Ref{Wqoexct} for the $V>0$ and $V<0$
phases agrees exactly with the fermion PSG's that we obtained above.  This
example shows that the PSG's introduced in \Ref{Wqoexct,Wqogen} 
are the symmetry groups of the hopping Hamiltonian of the ends of condensed
strings. The PSG description and the string-net-condensation description of
quantum orders are intimately related.

Here we would like to point out that the PSG's introduced in
\Ref{Wqoslpub,Wqogen} are all fermion PSG's. They are only one of many
different kinds of PSG's that can be used to characterize quantum orders.  In
general, a quantum ordered state may contain condensations of several types of
strings. The ends of each type of condensed strings will have their own PSG.

\section{Massless fermion and PSG in string-net condensed state}
\label{masslessferm}

In \Ref{Wqoslpub,WZqoind}, it was pointed out that PSG can protect
masslessness of the emerging fermions, just like symmetry can protect the
masslessness of Nambu-Goldstone bosons.  In this section, we are going to
study an exact soluble spin-$\frac12\frac12$ model with string-net
condensation and emerging massless fermions.  Through this soluble model, we
demonstrate how PSG that characterizes the string-net condensation can protect
the masslessness of the fermions.  The exact soluble model that we are going
to study is motivated by Kitaev's exact soluble spin-1/2 model on honeycomb
lattice.\cite{Ktoap}

\subsection{Exact soluble spin-$\frac12\frac12$ model}

The exact soluble model is a local bosonic model on square lattice.
To construct the model, we start with four Majorana fermions
$\la_{\v i}^a$, $a=x,\bar x,y,\bar y$ and one complex fermion $\psi$.
$\la_{\v i}^a$ satisfy
\begin{equation}
 \{ \la_{\v i}^a, \la_{\v j}^b \}=2\del_{ab}\del_{\v i\v j}
\end{equation}

We note that 
\begin{equation} 
\hat U_{\v i,\v i+{\v x}}=-i\la^{x}_{\v i}\la^{\bar x}_{\v i+{\v x}}, \ \ \
\hat U_{\v i,\v i+{\v y}}=-i\la^{y}_{\v i}\la^{\bar y}_{\v i+{\v y}} ,\ \ \
\hat U_{\v i\v j} =  \hat U_{\v j\v i}
\end{equation} 
form a commuting set of operators. Using such a commuting set of operators, we
can construct the following exact soluble interacting fermion model
\begin{align}
\label{Hfmassless}
 H =& g\sum_{\v i} \hat F_{\v i}
+t \sum_{\v i} (
i \hat U_{\v i,\v i+\v x}\psi_{\v i}^\dag\psi_{\v i+\v x}
+i \hat U_{\v i,\v i+\v y}\psi_{\v i}^\dag\psi_{\v i+\v y}
+h.c.)
\nonumber\\
\hat F_{\v i} =& 
\hat U_{\v i,\v i_1} 
\hat U_{\v i_1, \v i_2} 
\hat U_{\v i_2,\v i_3} 
\hat U_{\v i_3,\v i} 
\end{align}
where $\v i_1 = \v i+{\v x}$, $\v i_2 = \v i+{\v x}+{\v y}$,
$\v i_3 = \v i+{\v y}$, and $t$ is real.  
We will call $\hat F_{\v i}$ a $Z_2$ flux
operator.  To obtain the Hilbert space within which the Hamiltonian $H$
acts, we group $\la^{x,\bar x,y,\bar y}$ into two complex fermion
operators
\begin{equation}
2 \psi_{1,\v i} = \la^{x}_{\v i} +i\la^{\bar x}_{\v i}, \ \ \
2 \psi_{2,\v i} = \la^{y}_{\v i} +i\la^{\bar y}_{\v i}
\end{equation}
on each site.  The complex fermion operators $\psi_{1,2}$ and $\psi$ generate
an eight dimensional Hilbert space on each site.

Since $\hat U_{\v i\v j}$ commute with each other, we can find the common
eigenstates of the $\hat U_{\v i\v j}$ operators: $|\{s_{\v i\v j}\}, n\>$,
where $s_{\v i\v j}$ is the eigenvalue of $ \hat U_{\v i\v j}$, and $n$ labels
different degenerate common eigenstates . Since $( \hat U_{\v i\v j})^2=1$ and
$ \hat U_{\v i\v j}= \hat U_{\v j\v i}$ , $s_{\v i\v j}$ satisfies $s_{\v i\v
j}=\pm 1$ and $s_{\v i\v j}=s_{\v j\v i}$.  Within the subspace with a fixed
set of $s_{\v i\v j}$: $\{ |\{s_{\v i\v j}\}, n\> |n=1,2,...\}$, the
Hamiltonian has a form
\begin{align}
\label{Hsmassless}
 H =& g\sum_{\v i} f_{\v i}
+t \sum_{\v i} (
i  s_{\v i,\v i+\v x}\psi_{\v i}^\dag\psi_{\v i+\v x}
+i  s_{\v i,\v i+\v y}\psi_{\v i}^\dag\psi_{\v i+\v y}
+h.c.)
\nonumber\\
f_{\v i} =& 
s_{\v i,\v i_1} 
s_{\v i_1, \v i_2} 
s_{\v i_2,\v i_3} 
s_{\v i_3,\v i} 
\end{align}
which is a free fermion Hamiltonian. Thus we can find all the many-body
eigenstates of $\psi_{\v i}$: $|\{s_{\v i\v j}\},\Psi_n\>$ and their energies
$E(\{s_{\v i\v j}\},n)$ in each subspace.  This way we solve the interacting
fermion model exactly. 

We note that the Hamiltonian $H$ can only change the fermion number on each
site by an even number. Thus the $H$ acts within a subspace which has an even
number of fermions on each site. We will call the subspace physical Hilbert
space. The physical Hilbert space has only four states per site.  When defined
on the physical space, $H$ becomes a local bosonic system which actually
describes a spin $\frac12\times \frac12$ system (with no spin rotation
symmetry).  We will call such a system spin-$\frac12\frac12$ system.  To
obtain an expression of $H$ within the physical Hilbert space, we introduce
two Majorana fermions $\eta_{1,\v i}$ and $\eta_{2,\v i}$ to represent
$\psi_{\v i}$: $2\psi_{\v i}=\eta_{1,\v i} +i\eta_{2,\v i}$.  We note that
$\la^a\eta_1$, $a=x,\bar x,y,\bar y$, act within the  four dimensional
physical Hilbert space on each site, and thus are 4 by 4 matrices.  Also
$\{-i\la^a\eta_1,-i\la^b\eta_1\}=2\del_{ab}$, thus the four 4 by 4 matrices
$\la^a\eta_1$ satisfy the algebra of Dirac matrices.  Therefore we can express
$\la^a\eta_1$ in terms of Dirac matrices $\ga^a$:
\begin{align}
 \la^a\eta_1 =& i\ga^a  \nonumber\\
\ga^{x} = & \si^x\otimes \si^x , &
\ga^{\bar x} = & \si^y\otimes \si^x   \nonumber\\
\ga^{y} = & \si^z\otimes \si^x , & 
\ga^{\bar y} = & \si^0\otimes \si^y  
\end{align}
We can also define the $\ga^5$
\begin{align}
 \ga^5 \equiv & \ga^{x}\ga^{\bar x}\ga^{y}\ga^{\bar y}
= -\si^0\otimes \si^z  \nonumber\\
=& \la^{x}\la^{\bar x}\la^{y}\la^{\bar y} = i\eta_1\eta_2
\end{align}
where we have used $1-2\psi^\dag\psi= -i\eta_1\eta_2$ and $
(-i\la^{x}\la^{\bar x}) (-i\la^{y}\la^{\bar y}) (-i\eta_1\eta_2)=1$ for states
with even numbers of fermions.  With the above definition of $\ga^a$ and
$\ga^5$, we find that
\begin{equation}
 \la^a\eta_2 = \ga^a \ga^5
\end{equation}
and
\begin{align}
\label{gapma}
 \la^a\psi =& \frac{i}{2} (\ga^a+\ga^a\ga^5)\equiv i\ga^{-,a},\ \ \
\nonumber\\
 \la^a\psi^\dag =& \frac{i}{2} (\ga^a-\ga^a\ga^5)\equiv i\ga^{+,a}  
\nonumber\\
 \ga^{-,a}=&(\ga^{+,a})^\dag
\end{align}
We also have
\begin{equation}
\label{gaab}
 \la^a\la^b=\ga^a\ga^b\equiv \ga^{ab}
\end{equation}
The above relations allows us to write $H$ in terms of 4 by 4 Dirac matrices.
For example
\begin{equation}
\label{Fga}
 \hat F_{\v i}= -
\ga^{yx}_{\v i}
\ga^{\bar x y}_{\v i+\v x}
\ga^{\bar  y\bar x}_{\v i+\v x+\v y}
\ga^{x \bar y}_{\v i+\v y}
\end{equation}
and
\begin{align}
 \hat U_{\v i,\v i+\v x}\psi^\dag_{\v i}\psi_{\v i+\v x}
=&  -i \ga^{+,x}_{\v i}\ga^{-,\bar x}_{\v i+\v x}  \nonumber\\
 \hat U_{\v i,\v i+\v y}\psi^\dag_{\v i}\psi_{\v i+\v y}
=&  -i \ga^{+,y}_{\v i}\ga^{-,\bar y}_{\v i+\v y} 
\end{align}

The physical states in the physical Hilbert space are invariant under local
$Z_2$ gauge transformations generated by
\begin{align}
 G=&\prod_{\v i} G_{\v i}^{n_{\v i}} \nonumber\\
n_{\v i}=&{\psi^\dag_{1,\v i}\psi_{1,\v i}
+\psi^\dag_{2,\v i}\psi_{2,\v i}+ \psi^\dag_{\v i}\psi_{\v i}}
\end{align}
where $G_{\v i}$ is an arbitrary function with only two values $\pm 1$ and
$n_{\v i}$ the number of fermions on site $\v i$.  We
note that the $Z_2$ gauge transformations change $\psi_{I\v i} \to G_{\v
i}\psi_{I\v i}$. The projection into the physical Hilbert space with even
fermion per site makes our theory a $Z_2$ gauge theory.

Since the Hamiltonian $H$ in \Eq{Hfmassless} is $Z_2$ gauge invariant: $[G,
H]=0$, eigenstate of $H$ within the physical Hilbert space can be obtained
from $|\{s_{\v i\v j}\}, \Psi_n\>$ by projecting into the physical Hilbert
space: $ \cP |\{s_{\v i\v j}\}, \Psi_n\>$. The projected state $ \cP |\{s_{\v
i\v j}\}, \Psi_n\>$ (or the physical state), if non-zero, is an eigenstate of
the spin-$\frac12\frac12$ model with energy $E(\{ s_{\v i\v j}\}, n)$.
The $Z_2$ gauge invariance implies that
\begin{align}
 \cP |\{s_{\v i\v j}\}, \Psi_n\> =& \cP |\{\t s_{\v i\v j}\}, \Psi_n\>
\nonumber\\
E(\{ s_{\v i\v j}\}, n)=&E(\{ \t s_{\v i\v j}\}, n)
\end{align}
if $s_{\v i\v j}$ and $\t s_{\v i\v j}$ are $Z_2$ gauge equivalent
\begin{equation}
\label{Z2trans}
 \t s_{\v i\v j} =G(\v i) s_{\v i\v j} G(\v j).
\end{equation}

Let us count the states to show that the projected states $ \cP |\{s_{\v i\v
j}\}, \Psi_n\>$ generate all states in the physical Hilbert space.
Let us consider a periodic lattice with $N_{site}=L_xL_y$ sites.
First there are $2^{2N_{site}}$ choices of $s_{\v i\v j}$. We note that 
there are $2^{N_{site}}$ different $Z_2$ gauge transformations. But the
constant gauge transformation $G(\v i)=-1$ does not change
$s_{\v i\v j}$. Thus there are $2^{N_{site}}/2$ different $ s_{\v i\v j}$'s
in each $Z_2$ gauge equivalent class. Therefore, there are
$2\times2^{N_{site}}$ different $Z_2$ gauge equivalent classes of
$s_{\v i\v j}$'s. We also note that
\begin{align}
& \prod_{\v i} s_{\v i,\v i+\v x} s_{\v i,\v i+\v y}  \nonumber\\
=& (-)^{L_x+L_y}\prod_{\v i} (-i \la^{x}_{\v i} \la^{\bar x}_{\v i})
(-i\la^{y}_{\v i} \la^{\bar y}_{\v i})   \nonumber\\
=& (-)^{L_x+L_y+\sum_{\v i}
( \psi^\dag_{1,\v i}\psi_{1,\v i} +\psi^\dag_{2,\v i}\psi_{2,\v i}) }
\end{align}
Thus, among the $2\times2^{N_{site}}$ different classes of $s_{\v i\v j}$'s,
$2^{N_{site}}$ of them satisfy $\prod_{\v i} s_{\v i,\v i+\v x} s_{\v i,\v
i+\v y}=(-)^{L_x+L_y}$ and have even numbers of $\psi_{1,\v i}$ and
$\psi_{2,\v i}$ fermions.  The other $2^{N_{site}}$ of them satisfy $\prod_{\v
i} s_{\v i,\v i+\v x} s_{\v i,\v i+\v y}=-(-)^{L_x+L_y}$ and have odd numbers
of $\psi_{1,\v i}$ and $\psi_{2,\v i}$ fermions.

For each fixed $s_{\v i\v j}$, there are $2^{N_{site}}$ many-body states of
the $\psi_{\v i}$ fermions, \ie $n$ in $|\{s_{\v i\v j}\}, \Psi_n\>$ runs from
1 to $2^{N_{site}}$.  Among those $2^{N_{site}}$ many-body states,
$2^{N_{site}}/2$ of them have even numbers of $\psi_{\v i}$ fermions and
$2^{N_{site}}/2$ of them have odd numbers of $\psi_{\v i}$ fermions.  In order
for the projection $\cP|\{s_{\v i\v j}\}, \Psi_n\>$ to be non-zero, the total
number of fermions must be even.  A physical state has even numbers of
$(\psi_{1,\v i},\psi_{2,\v i})$ fermions and even numbers of $\psi_{\v i}$
fermions, or it has odd numbers of $(\psi_{1,\v i},\psi_{2,\v i})$
fermions and odd numbers of $\psi_{\v i}$ fermions.  Thus there are
$2^{N_{site}}\times 2^{N_{site}}/2 +2^{N_{site}}\times 2^{N_{site}}/2=
4^{N_{site}} $ distinct physical states that can be produced by the
projection.  Thus the projection produces all the states in the physical
Hilbert space.

\subsection{Physical properties of the spin-$\frac12\frac12$ model}

Let us define a closed-string operator to be
\begin{equation}
W(C_{close})= \hat U_{\v i_1\v i_2}\hat U_{\v i_2\v i_3}...\hat U_{\v i_n\v i_1}
\end{equation} 
where $C_{close}$ is an closed oriented string $C_{close}=\v i_1\to \v
i_2...\to \v i_n\to \v i_1$ formed by nearest neighbor links. 
Since $C_{close}$ can intersect with itself, $C_{close}$ can also be viewed as
a closed string-net. We will also call $W(C_{close})$ a closed-string-net
operator.

The closed-string-net operators act within the physical Hilbert space and
commute with the Hamiltonian \Eq{Hfmassless}.  Thus there is a string-net
condensation since $\<W(C_{close})\>=\pm 1$ in the ground state of
\Eq{Hfmassless}.  The above strings correspond to the T3 string discussed in
section \ref{3type}. Unlike the spin-1/2 model, we do not have condensed T1
and T2 closed strings in the spin-$\frac12\frac12$ model.

We can also define open-string operators that act within the physical
Hilbert space
\begin{align}
\label{WCopen}
W(C_{open})=& \la^a_{\v i_1}\hat U_{\v i_1\v i_2}\hat U_{\v i_2\v i_3}...
\hat U_{\v i_{n-1}\v i_n}\la^b_{\v i_n} \nonumber\\
\t W(C_{open})=& \psi^\dag_{\v i_1}\hat U_{\v i_1\v i_2}\hat U_{\v i_2\v i_3}...
\hat U_{\v i_{n-1}\v i_n}\psi_{\v i_n} 
\end{align}
where $C_{open}$ is an open oriented string $C_{open}=\v i_1\to \v i_2...\to
\v i_n$ formed by nearest neighbor links.  $W(C)$ correspond to the open T3
string defined in section \ref{3type}.  Just like the spin-1/2 model \Eq{HV},
the ends of such strings correspond to gapped fermions (if $|g|\gg
|t|$). The ends of $\t W$ strings only differ from the ends of $W$
strings by a local bosonic operator. Thus the ends of $\t W$ strings are also
fermions.

\begin{figure}
\centerline{
\includegraphics[width=2.0in]{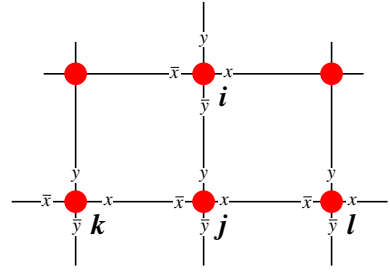}
}
\caption{
A particle can hop between different sites $\v i,\v j,\v k,\v l$.
}
\label{2Dhop}
\end{figure}

To really prove the ends of  $\t W$ strings are fermions, we need to show the
hopping of the ends of $\t W$ strings satisfy the fermion hopping algebra
introduced in \Ref{LW0360}:
\begin{align}
\label{fhopalg1}
t_{jl} t_{kj} t_{ji} =& - t_{ji}t_{kj} t_{jl}, \nonumber\\
[t_{ij},t_{kl}]= &0,\ \ \hbox{if $i,j,k,l$ are all different,}
\end{align}
where $t_{ji}$ describes the hopping from site $i$ to site $j$.  It was shown
that the particles are fermions if their hopping satisfy the algebra
\Eq{fhopalg1}.  We note that the ends of the $\t W$ strings live on the sites.
The labels $i,j,...$ in the above equation correspond to lattice sites $\v
i,\v j,...$.  The hops between sites $\v i,\v j,\v k,\v l$ in Fig. \ref{2Dhop}
are given by
\begin{align}
t_{\v i+\v a,\v i}=& \psi^\dag_{\v i+\v a}\hat U_{\v i+\v a,\v i}\psi_{\v i}
,\ \ \v a= \pm \v x,\pm \v y
\end{align}
Note that the hops between nearest neighbors are taken from the Hamiltonian
\Eq{Hfmassless}. 
Since $\hat U_{\v i\v j}$ commute with each other, the algebra of the
above hopping operators is just that of fermion hopping operators. In
particular, the above hopping operators satisfy the fermion hopping algebra
\Eq{fhopalg1}.  Hence, the ends of the $\t W$ strings are fermions.

For each fixed configuration $s_{\v i\v j}$, there are $2^{N_{site}}/2$
different states (with even or odd numbers of total $\psi$ fermions). Their
energy are given by the fermion hopping Hamiltonian \Eq{Hsmassless}.  Let
$E_0(\{s_{\v i\v j}\})$ be the ground state energy of \Eq{Hsmassless}.  The
ground state and the ground state energy of our spin-$\frac12\frac12$ model
\Eq{Hfhop} is obtained by choosing a configuration $s_{\v i\v j}$ that
minimize $E_0(\{s_{\v i\v j}\})$.  We note that $E_0(\{s_{\v i\v j}\})$ is
invariant under the $Z_2$ gauge transformation \Eq{Z2trans}. 

When $g\gg |t|$, the ground state of \Eq{Hfmassless} has $\hat F_{\v i}=-1$ which
minimize the dominating $g\sum_{\v i}\hat F_{\v i}$ term.  The ground state
configuration is given by
\begin{equation}
\label{spiflux}
 s_{\v i,\v i+\v x}=(-)^{i_y},\ \ \
 s_{\v i,\v i+\v y}=1.
\end{equation}
The $\psi_{\v i}$ fermion hopping Hamiltonian \Eq{Hsmassless} for the above
configuration describes fermion hopping with $\pi$-flux per plaquette.  
The fermion spectrum has a form
\begin{equation}
 E_{\v k} = \pm 2 \sqrt{t^2 \sin^2(k_x) + t^2 \sin^2(k_y)} .
\end{equation}
The low energy excitations of such a hopping Hamiltonian are described by two
two-component massless Dirac fermions in 2+1D.  We see that the ends of the
$\t W$ strings are massless Dirac fermions.

Our model also contain $Z_2$ gauge excitations. The $Z_2$ vortices are created
by flipping $\hat F_{\v i}=-1$ to $\hat F_{\v i}=1$ in some plaquettes.  The
$Z_2$ vortex behaves like a $\pi$-flux to the gapless fermions. Thus the
gapless fermions carry a unit $Z_2$ charge. The low energy effective theory of
our model is massless Dirac fermions coupled to a $Z_2$ gauge field.

\subsection{Projective symmetry and massless fermions}
\label{PSGmassless}

We know that symmetry breaking can produce and protect gapless Nambu-Goldstone
modes. 
In \Ref{Wqoslpub,WZqoind}, it was proposed that, in addition to symmetry
breaking, quantum order can also produce and protect gapless excitations.  The
gapless excitations produced and protected by quantum order can be gapless
gauge bosons and/or gapless fermions. In this paper we show that the quantum
orders discussed in \Ref{Wqoslpub,WZqoind} are due to string-net condensations.
Therefore, more precisely it is string-net condensations that produce and protect
gapless gauge bosons and/or gapless fermions.  The string-net condensations and
gapless excitations are connected in the following way.  Let us consider a
Hamiltonian that has a symmetry described by a symmetry group $SG$.  We assume
the ground state has a string-net condensation.  Then, the hopping Hamiltonian for
the ends of condensed string will be invariant under a larger group - the
projective symmetry group $PSG$, as discussed in section \ref{PSGend}.  $PSG$
is an extension of the symmetry group $SG$, \ie $PSG$ contain a normal
subgroup $IGG$ such that $PSG/IGG=SG$.  The relation between $PSG$ and gapless
gauge bosons is simple.  Let $\cG$ be the maximum continuous subgroup of
$IGG$. Then the gapless gauge bosons are described by a gauge theory with
$\cG$ as the gauge group.\cite{Wqoslpub,Wlight} Some times the ends of strings
are fermions.  However, the relation between gapless fermions and $PSG$ is
more complicated.  Through a case by case study of some
$PSG$'s\cite{Wqoslpub,WZqoind}, we find that certain $PSG$'s indeed guarantee
the existence of gapless fermions.

In this section, we are going to study a large family of exact soluble
local bosonic models which depends on many continuous parameters.
The ground states of the local bosonic models have a string-net condensation and
do not break any symmetry. We will show that the projective symmetry of the
ends of condensed strings protects a massless fermion. As a result, our exact
soluble model always has massless fermion excitations regardless the value of
the continuous parameters (as long as they are within a certain range).
This puts the results of \Ref{Wqoslpub,WZqoind}, which were based on mean-field
theory, on a firmer ground. 

The exact soluble local bosonic models are the spin-$\frac12\frac12$ model
\begin{align}
\label{Hfhop}
 H_{\frac12\frac12}=&
-g \sum_{\v i} \ga^{yx}_{\v i}
\ga^{\bar x y}_{\v i+\v x}
\ga^{\bar  y\bar x}_{\v i+\v x+\v y}
\ga^{x \bar y}_{\v i+\v y} \nonumber\\
&+\sum_{\v i} \Big(
 t \ga^{+,x}_{\v i}\ga^{-,\bar x}_{\v i+\v x}
+ t \ga^{+,y}_{\v i}\ga^{-,\bar y}_{\v i+\v y} +
h.c.\Big)
\end{align}
where $\ga^{ab}$ and $\ga^{\pm,a}$ are given in \Eq{gaab} and \Eq{gapma}.
We will discuss a more general Hamiltonian later.

The Hamiltonian is not invariant under $x\to -x$ parity $P_x$.  But it has a
$x\to -x$ parity symmetry if $P_x$ is followed by a spin rotation $\ga^x
\leftrightarrow \ga^{\bar x}$.  That is $\ga_{P_x}P_x H (\ga_{P_x}P_x)^{-1}=H$
with
\begin{equation}
 \ga_{P_x} = \ga^5 \frac{\ga^x-\ga^{\bar x}}{\sqrt{2}}
\end{equation}
Similarly for $y\to -y$ parity $P_y$, we have
$\ga_{P_y}P_y H (\ga_{P_y}P_y)^{-1}=H$ with
\begin{equation}
 \ga_{P_y} = \ga^5 \frac{\ga^y-\ga^{\bar y}}{\sqrt{2}}
\end{equation}

In the fermion representation $\ga_{P_x}$ and $\ga_{P_y}$
generate the following transformations
\begin{align}
 \ga_{P_x}:&\ \ 
\la^x_{\v i} \leftrightarrow \la^{\bar x}_{\v i},\ \ 
\psi_{\v i} \leftrightarrow \psi^\dag_{\v i},
\nonumber\\
 \ga_{P_y}:&\ \ 
\la^y_{\v i} \leftrightarrow \la^{\bar y}_{\v i},\ \ 
\psi_{\v i} \leftrightarrow \psi^\dag_{\v i}.
\end{align}

Now let us study how the symmetries $T_{x,y}$ and $\ga_{P_{x,y}}P_{x,y}$
are realized in the hopping Hamiltonian \Eq{Hsmassless} for the ends of
condensed strings.  As discussed in section \ref{PSGend}, the hopping
Hamiltonian may not be invariant under the symmetry transformations $T_{x,y}$
and $\ga_{P_{x,y}}P_{x,y}$ directly.  The hopping Hamiltonian only has a
projective symmetry generated by a symmetry transformation followed by a $Z_2$
gauge transformation $G(\v i)$. Since the $\pi$-flux configuration does not
break any symmetries, we expect the hopping Hamiltonian for the $\pi$-flux configuration to be invariant under 
$G_xT_x$,
$G_yT_y$,
$G_{P_x}\ga_{P_x}P_x$, and
$G_{P_y}\ga_{P_y}P_y$,
where $G_{x,y}$ and $G_{P_{x,y}}$  are the
corresponding gauge transformations.
The action of $T_{x,y}$ and $\ga_{P_{x,y}}P_{x,y}$ on the $\psi$ fermion
are given by
\begin{align}
 T_x:&\ \ \psi_{(i_x,i_y)} \to \psi_{(i_x+1,i_y)},\nonumber\\
 T_y:&\ \ \psi_{(i_x,i_y)} \to \psi_{(i_x,i_y+1)},\nonumber\\
\ga_{P_{x}} P_x:&\ \ \psi_{(i_x,i_y)} \leftrightarrow \psi^\dag_{(-i_x,i_y)},
\nonumber\\
\ga_{P_{y}} P_y:&\ \ \psi_{(i_x,i_y)} \leftrightarrow \psi^\dag_{(i_x,-i_y)}.
\end{align}
For the $\pi$-flux configuration \Eq{spiflux}, we need to choose the following
$G_{x,y}$ and $G_{P_{x,y}}$ in order for the combined transformation
$G_{x,y}T_{x,y}$ and $G_{P_{x,y}}\ga_{P_{x,y}}P_{x,y}$ to be the
symmetries of the hopping Hamiltonian \Eq{Hsmassless}
\begin{align}
 G_x&=1, & G_y=& (-)^{i_x},   \nonumber\\
 G_{P_x}&=(-)^{i_x}, & G_{P_y}=& (-)^{i_y}.
\end{align}
The hopping Hamiltonian is also invariant under a global $Z_2$ gauge
transformation
\begin{equation}
G_0:\ \ \psi_{\v i} \to -\psi_{\v i}
\end{equation}
The transformations $\{G_0,G_{x,y}T_{x,y},
G_{P_{x,y}}\ga_{P_{x,y}}P_{x,y}\}$ generate the PSG of the hopping
Hamiltonian. 

To show that the above PSG protects the masslessness of the fermions,
we consider a more general Hamiltonian by adding
\begin{equation}
 \del H_{\frac12\frac12} =\sum_{C_{\v i\v j}}
 \Big( t(C_{\v i\v j}) \t W(C_{\v i\v j}) +h.c \Big)
\end{equation}
to $H_{\frac12\frac12}$, where $C_{\v i\v j}$ is an open string connecting
site $\v i$ and site $\v j$ and $\t W(C_{\v i\v j})$ is given in \Eq{WCopen}.
The new Hamiltonian is still exactly soluble.  We will choose $t(C_{\v i\v
j})$ such that the new Hamiltonian has the translation symmetries and the
$P_{x,y}$ parity symmetries.  In the following, we would like to show that the
new Hamiltonian with those symmetries always has massless Dirac fermion
excitations (assuming $t(C_{\v i\v j})$ is not too big comparing to $g$).

When $t(C_{\v i\v j})$ is not too large, the ground state is still described by
the $\pi$-flux configuration. The new hopping Hamiltonian
for $\pi$-flux configuration has a more general form
\begin{equation}
 H=\sum_{\<\v i\v j\>} (\chi_{\v i\v j}\psi^\dag_{\v i}\psi_{\v j}+h.c.)
\end{equation}
The symmetry of the physical spin-$\frac12\frac12$ Hamiltonian requires
that the above hopping Hamiltonian to be invariant under the PSG discussed
above. Such an invariance will guarantee the existence of massless fermions.

The invariance under $G_xT_x$ and $G_yT_y$ require that
\begin{equation}
 \chi_{\v i,\v i+\v m}=(-)^{i_y m_x}\chi_{\v m}
\end{equation}
In the momentum space, 
\begin{align}
 \chi(\v k_1,\v k_2)\equiv&
N_{site}^{-1}\sum_{\v i\v j} e^{-i\v k_1\cdot \v i+i\v k_2\cdot \v j}
\chi_{\v i\v j}  \nonumber\\
=& \eps_0(\v k_2)\del_{\v k_1-\v k_2} 
+  \eps_1(\v k_2)\del_{\v k_1-\v k_2+\v Q_y}
\end{align}
where
\begin{align}
\eps_0(\v k)=& \sum_{m_x=even} e^{i\v k\cdot \v m} \chi_{\v m}, \nonumber\\
\eps_1(\v k)=& \sum_{m_x=odd} e^{i\v k\cdot \v m} \chi_{\v m}.
\end{align}
We note that  $\eps_0(\v k)$ and $\eps_1(\v k)$ are periodic function in the
Brillouin zone. They also satisfy
\begin{equation}
\label{epseps}
 \eps_0(\v k)= \eps_0(\v k+\v Q_x),\ \ \
 \eps_1(\v k)= -\eps_1(\v k+\v Q_x).
\end{equation}
where $\v Q_x=\pi\v x$ and $\v Q_y=\pi\v y$.
In the momentum space, we can rewrite $H$ as
\begin{equation}
 H=\sum_{\v k} \Psi^\dag_{\v k} \Ga(\v k) \Psi_{\v k}
\end{equation}
where $\Psi^T_{\v k}=(\psi_{\v k}, \psi_{\v k+\v Q_y})$.
The sum $\sum_{\v k}$ is over the reduced
Brillouin zone: $-\pi<k_x<\pi$ and $-\pi/2<k_y<\pi/2$.
$\Ga(\v k)$ has a form
\begin{equation}
 \Ga(\v k)=
\bpm
\eps_0(\v k) & \eps_1(\v k+\v Q_y)\\
 \eps_1(\v k) & \eps_0(\v k+\v Q_y)
\epm
\end{equation}

Note that the transformation$\ga_{P_x}$: $\psi\leftrightarrow \psi^\dag$
 changes $\sum \chi_{\v i\v j}\psi^\dag_{\v i}\psi_{\v j}$ to
$\sum \t \chi_{\v i\v j}\psi^\dag_{\v i}\psi_{\v j}$ with
$\t\chi_{\v i\v j}=-\chi_{\v j\v i}$. Thus
the invariance under $G_{P_x}\ga_{P_x}P_x$ requires that
\begin{equation}
-\chi_{P_x\v j,P_x\v i}= G_{P_x}(\v i)\chi_{\v i\v j} G_{P_x}(\v j)
\end{equation}
or
\begin{equation}
 \chi_{-P_x \v m}=-(-)^{m_xm_y+m_x}\chi_{\v m}
\end{equation}
In the momentum space, the above becomes
\begin{align}
\label{epsepsx}
 \eps_0(P_y \v k) =& -\eps_0(\v k)  \nonumber\\
 \eps_1(P_y \v k) =& \eps_1(\v k+\v Q_y)
\end{align}
Similarly, the invariance under $G_{P_y}\ga_{P_y}P_y$ requires that
\begin{equation}
-\chi_{P_y\v j,P_y\v i}= G_{P_y}(\v i)\chi_{\v i\v j} G_{P_y}(\v j)
\end{equation}
or
\begin{equation}
 \chi_{-P_y \v m}=-(-)^{m_xm_y+m_y}\chi_{\v m}
\end{equation}
In the momentum space
\begin{align}
\label{epsepsy}
 \eps_0(P_x \v k) =& -\eps_0(\v k+\v Q_y)  \nonumber\\
 \eps_1(P_x \v k) =& -\eps_1(\v k)
\end{align}

We see that the translation $T_{x,y}$ and $x\to -x$ parity $\ga_{P_x}P_x$
symmetries of the spin-$\frac12\frac12$ Hamiltonian require that $\eps_0(\v
k)=-\eps_0(P_y\v k)$ and hence $\eps_0(\v k)|_{k_y=0}=0$.  Similarly, the
translation $T_{x,y}$ and $y\to -y$ parity $\ga_{P_y}P_y$ symmetries require
that $\eps_1(\v k)|_{k_x=0}=0$.  Thus $T_{x,y}$ and $\ga_{P_{x,y}}P_{x,y}$
symmetries require that $\Ga(\v k)|_{\v k=0}=0$.  Using \Eq{epseps}, we find
that $\Ga(0)=0$ implies that $\Ga(\v Q_x)=0$.  The spin-$\frac12\frac12$
Hamiltonian \Eq{Hfhop} has (at least) two two-component massless Dirac
fermions if it has two translation $T_{x,y}$ and two parity
$\ga_{P_{x,y}}P_{x,y}$ symmetries.  We see that string-net condensation and
the associated projective symmetry produce and protect massless Dirac
fermions.

\section{Massless fermions and string-net condensation on cubic lattice}

The above calculation and the 2D model can be generalized to 
3D cubic lattice.
We introduce six Majorana fermions $\la^a_{\v i}$,
where $a=x,\bar x,y,\bar y,z,\bar z$. 
One set of commuting operators on square lattice has a form
\begin{align}
 \hat U_{\v i,\v i+\v x} = & -i\la^{x}_{\v i} \la^{\bar x}_{\v i+\v x} \nonumber\\
 \hat U_{\v i,\v i+\v y} = & -i\la^{y}_{\v i} \la^{\bar y}_{\v i+\v y} \nonumber\\
 \hat U_{\v i,\v i+\v z} = & -i\la^{z}_{\v i} \la^{\bar z}_{\v i+\v z} \nonumber\\
 \hat U_{\v i,\v j}^\dag = & \hat U_{\v j,\v i}
\end{align}
Using
$\hat U_{\v i,\v j}$ and a complex fermion $\psi_{\v i}$, 
we can construct an exact soluble interacting Hamiltonian
on cubic lattice
\begin{align}
\label{HF3D}
 H_{\frac12\frac12\frac12} =& g\sum_{\v p}  \hat F_{\v p}
+t\sum_{\v i}\sum_{\v a=\v x,\v y,\v z}
\Big( i \hat U_{\v i,\v i+\v a}\psi^\dag_{\v i}\psi_{\v i+\v a} +h.c. \Big),
\nonumber\\
 \hat F_{\v p} = &
 \hat U_{\v i_1, \v i_2} 
 \hat U_{\v i_2,\v i_3} 
 \hat U_{\v i_3,\v i_4} 
 \hat U_{\v i_4,\v i_1} 
\end{align}
where $\sum_{\v p}$ sum over all the square faces of the cubic lattice. 
$\v i_1$,
$\v i_2$,
$\v i_3$, and
$\v i_4$ label the four corners of the square $\v p$.
The Hilbert space of system is generated by complex fermion
operators $\psi_{\v i}$ and
\begin{align}
 2\psi_{1,\v i} =& \la^{x}_{\v i} + i\la^{\bar x}_{\v i} \nonumber\\
 2\psi_{2,\v i} =& \la^{y}_{\v i} + i\la^{\bar y}_{\v i} \nonumber\\
 2\psi_{3,\v i} =& \la^{z}_{\v i} + i\la^{\bar z}_{\v i}
\end{align}
and there are 16 states per site.

A physical Hilbert space is defined as a subspace with even numbers of
fermions per site. The physical Hilbert space has 8 states per site.  When
restricted in the physical Hilbert space,
$H_{\frac12\frac12\frac12}$ defines our
spin-$\frac12\frac12\frac12$ system, which is a \emph{local bosonic system}.

When $g\gg |t|$, our spin-$\frac12\frac12\frac12$ model has two four-component
massless Dirac fermions as its low lying excitations.  The model also has a
$Z_2$ gauge excitations and the massless Dirac fermions carry unit $Z_2$ gauge
charge.  Again, the model has a string-net condensation in its ground state.
Both the $Z_2$ gauge excitation and the massless fermion are produced and
protected by the string-net condensation and the associated PSG.

\section{Artificial light and artificial massless electron 
on cubic lattice }
\label{QED}

In this section, we are going to combine the above 3D model and the rotor
model discussed in \Ref{MS0204} and \Ref{Walight} to obtain a quasi-exact
soluble local bosonic model that contains massless Dirac fermions coupled to
massless $U(1)$ gauge bosons.

\subsection{3D rotor model and artificial light}

A rotor is described by an angular variable $\hat\th$. The angular momentum of
$\hat\th$, denoted as $S^z$, is quantized as integers.  The 3D rotor model under
consideration has one rotor on every link of a cubic lattice.  We use $\v i\v
j$ to label the nearest neighbor links.  $\v i\v j$ and $\v j\v i$
label the same links. For convenience, we
will define $\hat\th_{\v i\v j}=-\hat\th_{\v j\v i}$ and
$S^z_{\v i\v j}=-S^z_{\v j\v i}$.
The 3D rotor Hamiltonian has a form
\begin{align}
\label{Hrotorlatt3D}
 H_{rotor} & = U \sum_{\v i} 
\left(\sum_{\v a}
 S^z_{\v i,\v i+\v a} \right)^2 
+ \frac12 J \sum_{\v i, \v a} (S^z_{\v i,\v i+\v a})^2   \nonumber\\
&+g_1\sum_{\v p} 
\cos(
 \hat\th_{\v i_1\v i_2}
+\hat\th_{\v i_2\v i_3}
+\hat\th_{\v i_3\v i_4}
+\hat\th_{\v i_4\v i_1}
)
\end{align}
Here $\v i=(i_x,i_y,i_z)$ label the sites of the cubic lattice, and $\v a=\pm
\v x, \pm \v y, \pm \v z$.  The $\sum_{\v p}$ sum over all the square faces of
the cubic lattice.  $\v i_1$, $\v i_2$, $\v i_3$, and $\v i_4$ label the four
corners of the square $\v p$.

When $J=g_1=0$ and $U>0$, the state with all $S^z_{\v i\v j}=0$
is a ground state. Such a state will be regarded as a state with no strings.
We can create a string or a string-net from the no-string state using the
following string (or string-net) operator
\begin{equation}
 W_{U(1)}(C)=\prod_C e^{i\hat\th_{\v i\v j}}
\end{equation}
where $C$ is a string (or a string-net) formed by the nearest neighbor links,
and $\prod_C$ is a product over all the nearest neighbor links $\v i\v j$ on
the string (or string-net).  Since the closed-string-net operator
$W_{U(1)}(C_{close})$ commute with $H_{rotor}$ when $J=g_1=0$,
$W_{U(1)}(C_{close})$ generate a large set of degenerate ground states.  The
degenerate ground states are described by closed string-nets.

There is another way to generate the degenerate ground states. We note that
all the degenerate ground states satisfy $\sum_{\v a} S^z_{\v i,\v i+\v a}=0$.
Let $|\{\th_{\v i\v j} \}\>$ be the common eigenstate of $\hat\th_{\v i\v j}$:
$\hat\th_{\v i\v j} |\{\th_{\v i\v j} \}\>= \th_{\v i\v j}|\{\th_{\v i\v j}
\}\>$.  Then the projection into $\sum_{\v a} S^z_{\v i,\v i+\v a}=0$
subspace: $\cP |\{\th_{\v i\v j} \}\>$ give us a degenerate ground state.  We
note that 
\begin{equation}
 e^{i\sum_{\v i}\phi_{\v i}\sum_{\v a} S^z_{\v i,\v i+\v a}}
\end{equation}
generate a $U(1)$ gauge transformation
$|\{\th_{\v i\v j} \}\> \to |\{\t \th_{\v i\v j} \}\>$, where
\begin{equation}
 \t \th_{\v i\v j} = \th_{\v i\v j}+\phi_{\v i}-\phi_{\v j}
\end{equation}
Thus two $U(1)$ gauge equivalent configurations $\th_{\v i\v j}$ and
$\t \th_{\v i\v j}$ give rise to the same projected state
\begin{equation}
 \cP |\{\th_{\v i\v j} \}\>= \cP |\{\t\th_{\v i\v j} \}\>
\end{equation}
We find that the degenerate ground states are described by $U(1)$ gauge
equivalent classes of $\th_{\v i\v j}$. The degenerate ground states also have
a $U(1)$ gauge structure.

When $J=0$ but $g_1\neq 0$, the degeneracy in the ground states are lifted.
One can show that, in this case, $\cP |\{\th_{\v i\v j} \}\>$ is an energy
eigenstate with an energy $g_1\sum_{\v p}\cos( \th_{\v i_1\v i_2} +\th_{\v
i_2\v i_3} +\th_{\v i_3\v i_4} +\th_{\v i_4\v i_1}) $. Clearly two $U(1)$
gauge equivalent configurations $\th_{\v i\v j}$ and $\t \th_{\v i\v j}$ have
the same energy. The non-zero $g_1$ makes the closed string-nets to fluctuate
and vanishing $J$ means that the strings in the string-nets have no tension.
Thus $J=0$ ground state has strong fluctuations of large closed string-nets,
and the ground state has a closed-string-net condensation.\cite{Walight}

When $J\neq 0$, $\cP |\{\th_{\v i\v j} \}\>$ is no longer an eigenstate.
The fluctuations of $\th_{\v i\v j}$ describe a dynamical $U(1)$ gauge
theory with $\th_{\v i\v j}$ as the gauge potential.\cite{MS0204,Wlight}

\subsection{(Quasi-)exact soluble QED on cubic lattice}

To obtain massless Dirac fermions and $U(1)$ gauge bosons from a local bosonic
model, we mix the
spin-$\frac12\frac12\frac12$ model and the rotor model to get
\begin{align}
\label{3DQED}
 H_{QED} & = U \sum_{\v i} 
\left(\psi^\dag_{\v i}\psi_{\v i}+\sum_{\v a}
 S^z_{\v i,\v i+\v a} \right)^2 
+ \frac{J}{2}  \sum_{\v i, \v a} (S^z_{\v i,\v i+\v a})^2   \nonumber\\
&+g_1\sum_{\v p} \cos( \hat\Phi_{\v p} )  
+g\sum_{\v p}  \hat F_{\v p}
\\
& +t\sum_{\v i}\sum_{\v a=\v x,\v y,\v z}
\Big( i e^{i\hat\th_{\v i\v j}}
\hat U_{\v i,\v i+\v a}\psi^\dag_{\v i}\psi_{\v i+\v a} +h.c. \Big)
\nonumber 
\end{align}
where $\hat\Phi_{\v p} = \hat\th_{\v i_1\v i_2}
+\hat\th_{\v i_2\v i_3}
+\hat\th_{\v i_3\v i_4}
+\hat\th_{\v i_4\v i_1}$.
If we restrict ourselves within the physical Hilbert space with even numbers
of fermions per site, the above model is a local bosonic model.

Let us first set $J=0$. In this case, the above model can be solved exactly.
First let us also set $U=0$. In this case $\hat \th_{\v i\v j}$ and $\hat
U_{\v i\v j}$ commute with $H_{QED}$ and commute with each other. Let
$|\{\th_{\v i\v j}, s_{\v i\v j} \},n\>$ be the common eigenstates of $\hat
\th_{\v i\v j}$ and $\hat \hat U_{\v i\v j}$, where $n=1,2,...,2^{N_{site}}$
labels different degenerate common eigenstates.  Within the subspace expanded 
by $|\{\th_{\v i\v j}, s_{\v i\v j} \},n\>$ , $n=1,2,...,2^{N_{site}}$, the
$H_{QED}$ reduces to
\begin{align}
\label{thshop}
 H_{hop}  = 
&g_1\sum_{\v p} \cos( \Phi_{\v p} )  
+g\sum_{\v p}  f_{\v p}
\\
& +t\sum_{\v i}\sum_{\v a=\v x,\v y,\v z}
\Big( i e^{i\th_{\v i\v j}}
s_{\v i,\v i+\v a}\psi^\dag_{\v i}\psi_{\v i+\v a} +h.c. \Big) 
\nonumber 
\end{align}
which is a free fermion hopping model. Let $|\{\th_{\v i\v j}, s_{\v i\v j}
\},\Psi_n\>$ be the many-fermion eigenstate of the above fermion hopping model
and let $E(\{\th_{\v i\v j}, s_{\v i\v j} \}, n)$ be its energy. Then
$|\{\th_{\v i\v j)}, s_{\v i\v j} \},\Psi_n\>$ is also an eigenstate of
$H_{QED}|_{J=0,U=0}$ with energy $E(\{\th_{\v i\v j}, s_{\v i\v j} \}, n)$. 

We note that 
\begin{equation}
\hat N_{\v i}=
\psi^\dag_{\v i}\psi_{\v i}+\sum_{\v a}
 S^z_{\v i,\v i+\v a}
\end{equation}
commute with each other and commute with $H_{QED}$. Thus the
eigenstates of  $H_{QED}|_{J=0}$ can be obtained from the eigenstates
of $H_{QED}|_{J=0,U=0}$ by projecting into the subspace with
$\hat N_{\v i}=N_{\v i}$: 
\begin{equation}
\label{waveF}
 \cP_{\{N_{\v i}\}} |\{\th_{\v i\v j}, s_{\v i\v j} \},\Psi_n\>
\end{equation}
The above state is an eigenstate of $H_{QED}|_{J=0}$ with an energy
\begin{equation}
\label{UEthsn}
U\sum_{\v i} N_{\v i}+E(\{\th_{\v i\v j}, s_{\v i\v j} \}, n).
\end{equation}
\Eq{waveF} and \Eq{UEthsn} are our exact solution of $H_{QED}|_{J=0}$.  (We
have implicitly assumed that $\cP_{\{N_{\v i}\}}$ also perform the projection
into the physical Hilbert space of even numbers of fermions per site.)

When $U$ is positive and large, the low energy excitations only appear in the
sector $N_{\v i}=0$. Those low energy eigenstates are given by $\cP|\{\th_{\v
i\v j}, s_{\v i\v j} \},\Psi_n\> $ where $\cP$ is the projection into the
$N_{\v i}=0$ subspace and the even-fermion subspace. Their energy is
$E(\{\th_{\v i\v j}, s_{\v i\v j} \}, n)$.

Let us further assume that $-g_1\gg |t|$ and $g\gg |t|$.
In this limit, the ground state have $f_{\v p}=-1$
and $\Phi_{\v p}=0$. 
We can choose 
\begin{align}
\label{3Dconf}
 \th_{\v i,\v i+\v a}=&0,\ \ \v a=\v x,\v y,\v z  \nonumber\\
 s_{\v i,\v i+{\v x}} =& 1, \nonumber\\
 s_{\v i,\v i+{\v y}} =& (-)^{i_x} ,  \nonumber\\
 s_{\v i,\v i+{\v z}} =& (-)^{i_x+i_y} 
\end{align}
to describe such a configuration.  For such a configuration, \Eq{thshop}
describes a staggered fermion Hamiltonian.\cite{BSK7643,ID89,Z92} The ground
state wave function $\cP|\{\th_{\v i\v j}, s_{\v i\v j} \},\Psi_0\>$ is an
eigenstate of the $U(1)$ closed-string-net operator $W_{U(1)}(C_{close})$ with
eigenvalue $1$. It is also a eigenstate of the $Z_2$ closed-string-net
operator $W(C_{close})$ with eigenvalue $(-)^{N_p}$ where $N_p$ is the number
of the square plaquettes enclosed by $C_{close}$. We see that there is a
condensation of closed $U(1)$ and $Z_2$ string-nets in the $J=0$ ground state.
In such a string-net condensed state, there are gapless fermionic excitations,
which are described by fermion-hopping in $\pi$-flux phase.

In the momentum space, the fermion hopping Hamiltonian \Eq{thshop}
for the $\pi$-flux configuration has a form
\begin{align}
\label{thshopK}
 H_{hop}=& {\sum_{\v k}}^\prime 
\Psi^\dag_{a,\v k} \Ga(\v k) \Psi_{a,\v k} + \text{Const.}
\end{align}
where
\begin{align}
\Psi_{a,\v k}^T =& (
\psi_{a,\v k},
\psi_{a,\v k+\v Q_x},
\psi_{a,\v k+\v Q_y},
\psi_{a,\v k+\v Q_x+\v Q_y} 
),  \nonumber\\
\Ga(\v k) =& 
2t(\sin(k_x) \Ga_1 + \sin(k_y) \Ga_2 + \sin(k_z) \Ga_3 ) \nonumber 
\end{align}
and
$\Ga_1 = \tau^3\otimes \tau^0$,
$\Ga_2 = \tau^1\otimes \tau^3$, and
$\Ga_3 = \tau^1\otimes \tau^1$.
Here $\tau^{1,2,3}$ are the Pauli matrices and $\tau^0$ is the 2 by 2 identity
matrix.  The momentum summation $\sum_{\v k}^\prime$ is over a range $k_x \in
(-\pi/2,\pi/2)$, $k_y \in (-\pi/2,\pi/2)$, and $k_z \in (-\pi,\pi)$.  Since
$\{ \Ga_i, \Ga_j \} =2\del_{ij}$, $i,j=1,2,3$, we find the fermions have a
dispersion
\begin{align} 
E(\v k) = \pm 2t
\sqrt{ \sin^2(k_x) + \sin^2(k_y) + \sin^2(k_z)} 
\end{align} 
We see that the dispersion has two nodes at $\v k = 0$ and $\v k = (0,0,\pi)$.
Thus, \Eq{thshop} will give rise to 2 massless four-component Dirac
fermions in the continuum limit.

After including the $U(1)$ gauge fluctuations described by $\th_{\v i\v j}$
and the $Z_2$ gauge fluctuations described by $s_{\v i\v j}$, the massless
Dirac fermions interact with the $U(1)$ and the $Z_2$ gauge fields as fermions
with unit charge. Therefore the total low energy effective theory of our model
is a QED with 2 families of Dirac fermions of unit charge (plus an extra $Z_2$
gauge filed).  We will call those fermions artificial electrons.  The
continuum effective theory has a form
\begin{align}
\label{psiU1c}
 \cL =&
 \bar \psi_{I} D_0 \ga^0 \psi_{I} + 
 v_f\bar \psi_{I} D_i \ga^i \psi_{I} 
\nonumber\\
 &+ \frac{C}{Jl_0} \v E^2 - l_0g_1 \v B^2 +...
\end{align}
where $l_0$ is the lattice constant, $I=1,2$,
$D_0 =\prt_t + i a_0$, $D_i =\prt_i + i a_i|_{i=1,2,3}$, 
$v_f= 2l_0 t$, $\ga^\mu|_{\mu=0,1,2,3}$ are $4\times 4$ Dirac
matrices, and $\bar \psi_{I}=\psi_{I}^\dag \ga^0$. 

We like to pointed out the constant $C$ is \Eq{psiU1c} is of order $1$. Thus
the coefficient of the $\v E^2$ term $\frac{C}{Jl_0}\to \infty$ when $J=0$.
For a finite $J$, the $U(1)$ gauge field will have a non-trivial dynamics.  We
also like to point out that, without fine tuning, the speed of artificial
light, $c_a\sim l_0\sqrt{Jg_1}$, and the speed of artificial electrons, $v_f$,
do not have to be the same in our model.  Thus the Lorentz symmetry is not
guaranteed.

We would like to remark that, for finite $J$,
the $U(1)$ closed-string operators no longer condense. A necessary (but not
sufficient) condition for closed strings to condense is that the ground state
expectation value of the closed-string operator satisfy the perimeter law
\begin{equation}
 \<W_{U(1)}(C_{close})\>=A e^{-L_C/\xi}
\end{equation}
where $L_C$ is the length of the closed string and $(A,\xi)$  are constants
for large closed strings. We note that the closed-string operators are the
Wilson loop operator of the $U(1)$ gauge field. If the 3+1D $U(1)$ gauge
theory is in the Coulomb phase where the artificial light is gapless, it was
found that\cite{K7959}
\begin{equation}
 \<W(C_{close})\>=A(C) e^{-L_C/\xi}
\end{equation}
where $A(C)$ depend on the shape of the closed string $C_{close}$ even in the
large-string limit. Thus the closed strings in our model do not exactly
condense. The $U(1)$ Coulomb phase is, in some sense, similar to the
algebraic-long-range order phase of 1+1D interacting boson model where the
bosons do not exactly condense but the boson operator has an
algebraic-long-range correlation.

\subsection{Emerging chiral symmetry from PSG}

\Eq{psiU1c} describes the low energy dynamics of the ends of open strings (the
fermion $\psi$) and the ``condensed'' closed string-nets (the $U(1)$ gauge
field).  The fermions and gauge boson are massless and interact with each
other. Here we would like to address an important question: after integrating
out high energy fermions and gauge fluctuations, do fermions and gauge boson
remain to be massless? In general, interaction between massless excitations
will generate a mass term for them, unless the masslessness is protected by
symmetry or some other things.  We know that due to the $U(1)$ gauge
invariance, the radiative corrections cannot generate counter term that break
the $U(1)$ gauge invariance. Thus radiative corrections cannot generate mass
for the $U(1)$ gauge boson.  For the fermions, if the theory has a chiral
symmetry $\psi_{I}\to e^{i\th \ga^5}\psi_{I}$, $\ga^5= \ga^0\ga^1\ga^2\ga^3$,
then the radiative corrections cannot generate counter terms that break the
chiral symmetry and thus cannot generate mass for fermions.  Although the low
energy effective theory \Eq{psiU1c} appears to have the chiral symmetry, in
fact it does not. This is because that \Eq{psiU1c} is derived from a lattice
model. It contains many other higher order terms summarized by the $...$ in
\Eq{psiU1c}. Those higher order terms do not have the chiral symmetry. To see
this, we note that the action of $\ga^5$ on $\Psi_{a,\v k}$ is realized by a 4
by 4 matrix $\ga^5\propto \Ga_1\Ga_2\Ga_3 \propto \tau^3\otimes \tau^2$. We
also note that the periodic boundary conditions of $\Psi_{a,\v k}$ in the
reduced Brillouin zone are given by
\begin{equation}
 \Psi_{a,\v k+\v Q_x} = \tau^1\otimes \tau^0 \Psi_{a,\v k},\ \ 
 \Psi_{a,\v k+\v Q_y} = \tau^0\otimes \tau^1 \Psi_{a,\v k},\ \ 
\end{equation}
We find that the action of $\ga^5$ is incompatible with the periodic boundary
conditions since $\ga^5$ does not commute with $\tau^1\otimes \tau^0$ and
$\tau^0\otimes \tau^1$. Therefore the chiral symmetry generated by $\ga^5$
cannot be realized on lattice.  Due to the lack of chiral symmetry, it appears
that the radiative corrections can generate a mass term
\begin{equation}
\del \cL=  \bar \psi_{I, a}  m \psi_{I, a} 
\end{equation}
which is allowed by the symmetry.

The lack of chiral symmetry on lattice makes it very difficult to study
massless fermions/quarks in lattice gauge theory. In last a few years, this
problem was solved using the Ginsparg-Wilson
relation.\cite{GW8249,N9741,H9801,L9842} In the following, we would like to
show that there is another way to solve the massless-fermion/chiral-symmetry
problem. We will show that our model has an emerging chiral symmetry that
appear only at low energies. The low energy chiral symmetry comes from the
non-trivial quantum order and the associated PSG in the string-net condensed
ground state.\cite{Wqoslpub,Wlight,WZqoind}  
The Dirac operator in our model satisfies a linear relation
\begin{equation}
 WDW^\dag = D,\ \ \ W\in PSG
\end{equation}
in contrast to the non-linear Ginsparg-Wilson relation
\begin{equation}
 D\ga^5+\ga^5D=a D\ga^5 D
\end{equation}
Because of the low energy chiral symmetry, all the $2$ families
of Dirac fermions remain massless even after we include the radiative
corrections from the interaction with the $U(1)$ gauge bosons.

To see how the string-net condensation and the related PSG protect the massless
fermions, we follow closely the discussion in section \ref{PSGmassless}.
The Hamiltonian \Eq{3DQED} is a mixture of the rotor model and the
spin-$\frac12\frac12\frac12$ model.
The symmetry properties of the rotor part is simple. Here, we will concentrate
on the spin-$\frac12\frac12\frac12$ part.
\Eq{3DQED}
is not invariant under the six parity transformations:
$P_{x,y,z}$ and $P_{xy,yz,zx}$ that generate 
$x\leftrightarrow -x$,
$y\leftrightarrow -y$,
$z\leftrightarrow -z$,
$x\leftrightarrow y$,
$y\leftrightarrow z$, and
$z\leftrightarrow x$.
But it is invariant under parity $P_{x,y,z}$ and $P_{xy,yz,zx}$ 
followed by spin rotations
$\ga_{P_{x,y,z}}$  and
$\ga_{P_{xy,yz,zx}}$ respectively.
In the fermion representation $\ga_{P_{x,y,z}}$ and
$\ga_{P_{xy,yz,zx}}$ 
generate the following transformations
\begin{align}
 \ga_{P_x}:&\ \ 
\la^x_{\v i} \leftrightarrow \la^{\bar x}_{\v i},\ \ 
\psi_{\v i} \leftrightarrow \psi^\dag_{\v i},
\nonumber\\
 \ga_{P_y}:&\ \ 
\la^y_{\v i} \leftrightarrow \la^{\bar y}_{\v i},\ \ 
\psi_{\v i} \leftrightarrow \psi^\dag_{\v i},
\nonumber\\
 \ga_{P_z}:&\ \ 
\la^z_{\v i} \leftrightarrow \la^{\bar z}_{\v i},\ \ 
\psi_{\v i} \leftrightarrow \psi^\dag_{\v i},
\nonumber\\
 \ga_{P_{xy}}:&\ \ 
\la^x_{\v i} \leftrightarrow \la^y_{\v i},\ \ 
\la^{\bar x}_{\v i} \leftrightarrow \la^{\bar y}_{\v i},\ \ 
\nonumber\\
 \ga_{P_{yz}}:&\ \ 
\la^y_{\v i} \leftrightarrow \la^z_{\v i},\ \ 
\la^{\bar y}_{\v i} \leftrightarrow \la^{\bar z}_{\v i},\ \ 
\nonumber\\
 \ga_{P_{zx}}:&\ \ 
\la^z_{\v i} \leftrightarrow \la^x_{\v i},\ \ 
\la^{\bar z}_{\v i} \leftrightarrow \la^{\bar x}_{\v i},\ \ 
\end{align}

The symmetries $T_{x,y}$, $\ga_{P_{x,y,z}}P_{x,y,z}$, and
$\ga_{P_{xy,yz,zx}}P_{xy,yz,zx}$ are realized in the hopping Hamiltonian
\Eq{thshop} through PSG.  The hopping Hamiltonian is invariant only under the
symmetry transformations followed by proper $Z_2$ gauge transformations $G(\v
i)$.  Since the $\pi$-flux configuration $s_{\v i\v j}$ of the
spin-$\frac12\frac12\frac12$ sector and the zero-flux configuration $\th_{\v
i\v j}$ of the rotor sector do not break any symmetries, we expect the hopping
Hamiltonian \Eq{thshop} to be invariant under $G_{x,y,z}T_{x,y,z}$,
$G_{P_{x,y,z}}\ga_{P_{x,y,z}}P_{x,y,z}$, and
$G_{P_{xy,yz,zx}}\ga_{P_{xy,yz,zx}}P_{xy,yz,zx}$.  The action of $T_{x,y}$
and $\ga_{P_{xy,yz,zx}}P_{xy,yz,zx}$ on the $\psi$ fermion are standard
coordinate transformations.  The action of $\ga_{P_{x,y,z}}P_{x,y,z}$ on the
$\psi$ fermion are given by
\begin{align}
\ga_{P_{x}} P_x:&\ \ \psi_{(i_x,i_y,i_z)} \leftrightarrow 
\psi^\dag_{(-i_x,i_y,i_z)},
\nonumber\\
\ga_{P_{y}} P_y:&\ \ \psi_{(i_x,i_y,i_z)} \leftrightarrow 
\psi^\dag_{(i_x,-i_y,i_z)}.
\nonumber\\
\ga_{P_{y}} P_y:&\ \ \psi_{(i_x,i_y,i_z)} \leftrightarrow 
\psi^\dag_{(i_x,i_y,-i_z)}.
\end{align}
For the $\pi$-flux configuration \Eq{3Dconf}, we need to choose the following
$G_{x,y,z}$, $G_{P_{x,y,z}}$, and $G_{P_{xy,yz,zx}}$ in order for the combined
transformation $G_{x,y}T_{x,y}$, $G_{P_{x,y}}\ga_{P_{x,y}}P_{x,y}$, and
$G_{P_{xy,yz,zx}}\ga_{P_{xy,yz,zx}}P_{xy,yz,zx}$ to be the symmetries of the
hopping Hamiltonian \Eq{thshop}
\begin{align}
 G_x=&(-)^{i_y+i_z}, & G_y=& (-)^{i_z}, & G_z=& 1  \\
 G_{P_x}=&(-)^{i_x}, & G_{P_y}=& (-)^{i_y},  & G_{P_z}=& (-)^{i_z},
   \nonumber\\
 G_{P_{xy}}=&(-)^{i_xi_y}, &
 G_{P_{yz}}=&(-)^{i_yi_z}, &
 G_{P_{zx}}=&(-)^{i_xi_y+i_yi_z+i_zi_x} .
\nonumber 
\end{align}
The hopping Hamiltonian is also invariant under a global $Z_2$ gauge
transformation
\begin{equation}
G_0:\ \ \psi_{\v i} \to -\psi_{\v i}
\end{equation}
The transformations $\{G_{x,y,z}T_{x,y,z}$,
$G_{P_{x,y,z}}\ga_{P_{x,y,z}}P_{x,y,z}$,
$G_{P_{xy,yz,zx}}\ga_{P_{xy,yz,zx}}P_{xy,yz,zx}, G_0\}$ generate a PSG (a part
of the full PSG) of the hopping Hamiltonian.  

To study the robustness of massless fermions,
we consider a more general Hamiltonian by adding
\begin{equation}
 \del H =\sum_{C_{\v i\v j}}
 \Big( t(C_{\v i\v j}) \t W_{U(1)}(C_{\v i\v j}) +h.c \Big)
\end{equation}
to $H_{QED}$, where $C_{\v i\v j}$ is an open string
connecting site $\v i$ and site $\v j$ and $\t W_{U(1)}(C_{\v i\v j})$ 
an open-string operator
\begin{align}
\label{WU1open}
\t W_{U(1)}(C_{open})=& 
\psi^\dag_{\v i_1}e^{i\th_{\v i_1\v i_2}}\hat U_{\v i_1\v i_2}
...
e^{i\th_{\v i_{n-1}\v i_n}}\hat U_{\v i_{n-1}\v i_n}\psi_{\v i_n} 
\end{align}
The new Hamiltonian is still exactly soluble, when $J=0$.  We will choose
$t(C_{\v i\v j})$ such that the new Hamiltonian has the translation symmetries
and the $P_{x,y,z}$ parity symmetries.  We find that the resulting projective
symmetry imposes enough constraint on the hopping Hamiltonian for the ends of
condensed strings such that the Hamiltonian always has massless Dirac fermions
(assuming $t(C_{\v i\v j})$ is not too big comparing to $g$ and $g_1$).
Despite the PSG transformations
$G_{P_{xy,yz,zx}}\ga_{P_{xy,yz,zx}}P_{xy,yz,zx}$ are not needed for the
existence of the massless fermions, we will still include them in the
following discussion.

For small $t(C_{\v i\v j})$, the ground state is still described by
the $\pi$-flux configuration. The new hopping Hamiltonian
for $\pi$-flux configuration has a more general form
\begin{equation}
 H=\sum_{\<\v i\v j\>} (\chi_{\v i\v j}\psi^\dag_{\v i}\psi_{\v j}+h.c.)
\end{equation}
The symmetry of the generalized $H_{QED}$ requires that the above hopping
Hamiltonian to be invariant under the PSG generated by
$\{G_0,G_{x,y,z}T_{x,y,z}$, $
G_{P_{x,y,z}}\ga_{P_{x,y,z}}P_{x,y,z}$, $
G_{P_{xy,yz,zx}}\ga_{P_{xy,yz,zx}}P_{xy,yz,zx}\}$.

The invariance under $G_{x,y,z}T_{x,y,z}$ require that
\begin{equation}
 \chi_{\v i,\v i+\v m}=(-)^{i_y m_z} (-)^{i_x (m_y+m_z)} \chi_{\v m}
\end{equation}
In the momentum space, 
\begin{align}
 \chi(\v k_1,\v k_2)\equiv&
N_{site}^{-1}\sum_{\v i\v j} e^{-i\v k_1\cdot \v i+i\v k_2\cdot \v j}
\chi_{\v i\v j}  \nonumber\\
=& \sum_{\al,\bt=0,1}\eps_{\al\bt}(\v k_2)\del_{\v k_1-\v k_2+\al\v Q_x+
\bt\v Q_y} 
\end{align}
where
\begin{align}
\eps_{00}(\v k)=& \sum_{m_y+m_z=even,m_z=even} 
e^{i\v k\cdot \v m} \chi_{\v m}, \nonumber\\
\eps_{10}(\v k)=& \sum_{m_y+m_z=odd,m_z=even} 
e^{i\v k\cdot \v m} \chi_{\v m}, \nonumber\\
\eps_{01}(\v k)=& \sum_{m_y+m_z=even,m_z=odd} 
e^{i\v k\cdot \v m} \chi_{\v m}, \nonumber\\
\eps_{11}(\v k)=& \sum_{m_y+m_z=odd,m_z=odd} 
e^{i\v k\cdot \v m} \chi_{\v m}.
\end{align}
We note that  $\eps_{\al\bt}(\v k)$ are periodic function in the
lattice Brillouin zone $-\pi<k_{x,y,z}<\pi$. 
They also satisfy
\begin{equation}
\label{3Depseps}
 \eps_{\al\bt}(\v k)= (-)^\al\eps_{\al\bt}(\v k+\v Q_y+\v Q_z),\ \ \
 \eps_{\al\bt}(\v k)= (-)^\bt\eps_{\al\bt}(\v k+\v Q_z).
\end{equation}
The $\Ga(\v k)$ in \Eq{thshopK} now has a form
\begin{align}
& \Ga(\v k)=\\
&\bpm
\eps_{00}(\v k)&\eps_{10}(\v k+\v Q_x)&\eps_{01}(\v k+\v Q_y)
    &\eps_{11}(\v k+\v Q_x+\v Q_y)\\
\eps_{10}(\v k)&\eps_{00}(\v k+\v Q_x)&\eps_{11}(\v k+\v Q_y) 
    &\eps_{01}(\v k+\v Q_x+\v Q_y)  \\
\eps_{01}(\v k)&\eps_{11}(\v k+\v Q_x)&\eps_{00}(\v k+\v Q_y)
    &\eps_{10}(\v k+\v Q_x+\v Q_y)    \\
\eps_{11}(\v k)&\eps_{01}(\v k+\v Q_x)&\eps_{10}(\v k+\v Q_y) 
    &\eps_{00}(\v k+\v Q_x+\v Q_y)      
\epm
\nonumber 
\end{align}

Just as discussed in section \ref{PSGmassless},
the invariance under $G_{P_x}\ga_{P_x}P_x$ requires that
\begin{equation}
-\chi_{P_x\v j,P_x\v i}= G_{P_x}(\v i)\chi_{\v i\v j} G_{P_x}(\v j)
\end{equation}
or
\begin{equation}
 \chi_{-P_x \v m}=-(-)^{m_xm_y+m_ym_z+m_zm_x}(-)^{m_x} \chi_{\v m}
\end{equation}
In the momentum space, the above becomes
\begin{align}
\label{3Depsepsx}
 \eps_{00}(-P_x \v k) =& -\eps_{00}(\v k+\v Q_x),  \nonumber\\
 \eps_{10}(-P_x \v k) =& -\eps_{10}(\v k), \nonumber\\
 \eps_{01}(-P_x \v k) =& \eps_{01}(\v k+\v Q_x), \nonumber\\
 \eps_{11}(-P_x \v k) =& -\eps_{11}(\v k) .
\end{align}
where $\v Q_z=\pi\v z$.
Similarly, the invariance under $G_{P_y}\ga_{P_y}P_y$ requires that
\begin{equation}
 \chi_{-P_y \v m}=-(-)^{m_xm_y+m_ym_z+m_zm_x}(-)^{m_y} \chi_{\v m}
\end{equation}
In the momentum space
\begin{align}
\label{3Depsepsy}
 \eps_{00}(-P_y \v k) =& -\eps_{00}(\v k),  \nonumber\\
 \eps_{10}(-P_y \v k) =& \eps_{10}(\v k+\v Q_x), \nonumber\\
 \eps_{01}(-P_y \v k) =& -\eps_{01}(\v k), \nonumber\\
 \eps_{11}(-P_y \v k) =& -\eps_{11}(\v k+\v Q_x)  .
\end{align}
The invariance under $G_{P_z}\ga_{P_z}P_z$ requires that
\begin{equation}
 \chi_{-P_z \v m}=-(-)^{m_xm_y+m_ym_z+m_zm_x}(-)^{m_z} \chi_{\v m}
\end{equation}
In the momentum space
\begin{align}
\label{3Depsepsz}
 \eps_{00}(-P_z \v k) =& -\eps_{00}(\v k),  \nonumber\\
 \eps_{10}(-P_z \v k) =& -\eps_{10}(\v k+\v Q_x), \nonumber\\
 \eps_{01}(-P_z \v k) =& -\eps_{01}(\v k), \nonumber\\
 \eps_{11}(-P_z \v k) =& \eps_{11}(\v k +\v Q_x)  .
\end{align}

The invariance under $G_{P_{xy}}\ga_{P_{xy}}P_{xy}$ requires that
\begin{equation}
\chi_{P_{xy}\v i,P_{xy}\v j}= G_{P_{xy}}(\v i)\chi_{\v i\v j} G_{P_{xy}}(\v j)
\end{equation}
or
\begin{equation}
 \chi_{P_{xy} \v m}=(-)^{m_xm_y}\chi_{\v m}
\end{equation}
In the momentum space
\begin{align}
\label{3Depsepsxy}
 \eps_{00}(P_{xy} \v k) =& \eps_{00}(\v k),  \nonumber\\
 \eps_{10}(P_{xy} \v k) =& \eps_{10}(\v k+\v Q_x), \nonumber\\
 \eps_{01}(P_{xy} \v k) =& \eps_{01}(\v k+\v Q_x), \nonumber\\
 \eps_{11}(P_{xy} \v k) =& \eps_{11}(\v k)  .
\end{align}
The invariance under $G_{P_{yz}}\ga_{P_{yz}}P_{yz}$ requires that
\begin{equation}
 \chi_{P_{yz} \v m}=(-)^{m_ym_z}\chi_{\v m}
\end{equation}
or
\begin{align}
\label{3Depsepsyz}
 \eps_{00}(P_{yz} \v k) =& \eps_{00}(\v k),  \nonumber\\
 \eps_{10}(P_{yz} \v k) =& -\eps_{10}(\v k), \nonumber\\
 \eps_{01}(P_{yz} \v k) =& -\eps_{01}(\v k), \nonumber\\
 \eps_{11}(P_{yz} \v k) =& \eps_{11}(\v k)  .
\end{align}
The invariance under $G_{P_{zx}}\ga_{P_{zx}}P_{zx}$ requires that
\begin{equation}
 \chi_{P_{zx} \v m}=(-)^{m_xm_y+m_ym_z+m_zm_x}\chi_{\v m}
\end{equation}
or
\begin{align}
\label{3Depsepszx}
 \eps_{00}(P_{zx} \v k) =& \eps_{00}(\v k),  \nonumber\\
 \eps_{10}(P_{zx} \v k) =& \eps_{10}(\v k+\v Q_x), \nonumber\\
 \eps_{01}(P_{zx} \v k) =& -\eps_{01}(\v k), \nonumber\\
 \eps_{11}(P_{zx} \v k) =& \eps_{11}(\v k+\v Q_x)  .
\end{align}

We see that \Eq{3Depsepsx} require that $\eps_{10}(\v k)|_{k_y=k_z=0}=0$ and
$\eps_{11}(\v k)|_{k_y=k_z=0}=0$.  \Eq{3Depsepsz} require that $\eps_{00}(\v
k)|_{k_x=k_y=0}=0$ and $\eps_{01}(\v k)|_{k_x=k_y=0}=0$.  Thus the
$\eps_{\al\bt}(0)=0$. When combined with \Eq{3Depseps}, \Eq{3Depsepsx}, and
\Eq{3Depsepsz}, we find
\begin{equation}
 \eps_{\al\bt}( \al_x\v Q_x+ \al_y\v Q_y+ \al_z\v Q_z)=0,\ \ \ 
 \al_x,\al_y,\al_z=0,1
\end{equation}
Therefore $\Ga(\v k)=0$ when $\v k=0,\v Q_z$.  The two translation $T_{x,y}$
and the three parity $\ga_{P_{x,y,z}}P_{x,y,z}$ symmetries of $H_{QED}$
guarantee the existence of at least 2 four-component massless Dirac fermions.
Or more precisely, no symmetric local perturbations in the local bosonic model
$H_{QED}$ can generate mass terms for the two massless Dirac fermions in the
unperturbed Hamiltonian.

Since the mass term in the continuum effective field theory is not allowed by
the underlying lattice PSG, we say that our model has an emerging chiral
symmetry. The masslessness of the Dirac fermion is protected by the quantum
order and the associated PSG.

\section{
QED and QCD 
from a bosonic model on cubic lattice}
\label{secQCD}

In this section, we are going to generalize the results in \Ref{RS8909} and
\Ref{Wlight} and use a bosonic model on cubic lattice to generate QED and QCD
with $2N_f$ families of massless quarks and leptons.
To describe the local Hilbert space on site $\v i$ in our bosonic model, it is
convenient to introduce fermions $\la^a_{\v i}$ and $\psi^{n\al}_{\v i}$,
where $a=1,...,N_f$, $n=1,...,2N_f$ 
and $\al=1,2,3$.  $\la^a_{\v i}$ is
in the fundamental representation of a $SU(N_f)$ group.
$\psi^{n\al}_{\v i}$ is in the fundamental representation of a $SU(3)$
color group and a $SU(2N_f)$ group.  The Hilbert space of fermions
is bigger than the Hilbert space of our boson model. Only a physical subspace
of the fermions Hilbert space becomes the Hilbert space of our boson model.
The physical states on each site is formed by color singlet states 
that satisfy  
\begin{equation}
 \left(\la^{a\dag}_{\v i}\la^a_{\v i}\del^{\al\bt}
+\psi^{n\al \dag}_{\v i}\psi^{n\bt}_{\v i} 
-\del^{\al\bt}\frac{3}{2}N_f \right) \ket{\Phi_{phys}} =0
\end{equation}
where $N_f$ is assumed to be even.  
Once restricted within the physical
Hilbert space, the fermion model becomes our local bosonic model.

In the fermion representation,
the local physical operators in our bosonic model are given by
\begin{align}
 S^{mn}_{\v i} = &
\psi^{m\al \dag}_{\v i}\psi^{n\al}_{\v i}
-\frac{1}{2N_f}\del^{mn} \psi^{l\al \dag}_{\v i}\psi^{l\al}_{\v i}
\nonumber\\
 M^{ab}_{\v i} = &
\la^{a\dag}_{\v i}\la^{b}_{\v i}
-\frac{1}{N_f}\del^{ab} \la^{c\dag}_{\v i}\la^{c}_{\v i}
\nonumber\\
 \Ga^{a,lmn}_{\v i} = &
\la^{a\dag}_{\v i}
\psi^{l\al}_{\v i}
\psi^{m\bt}_{\v i}
\psi^{n\ga}_{\v i}\eps_{\al\bt\ga}
\end{align}
We note that by definition $M^{aa}_{\v i}=S^{nn}_{\v i}=0$.
The Hamiltonian of our boson model is given by
\begin{align}
\label{SUNfspin}
 H =& 
\frac{J_1}{N_f} \sum_{\<\v i\v j\>} S^{mn}_{\v i}S^{nm}_{\v j} 
+\frac{J_2}{N_f} \sum_{\<\v i\v j\>} M^{ab}_{\v i}M^{ba}_{\v j} 
\nonumber\\
&+\frac{J_3}{N_f^3} \sum_{\<\v i\v j\>} 
[\Ga^{a,lmn}_{\v i}\Ga^{a,lmn\dag}_{\v j} + h.c.]
\end{align}
Let us assume, for the time being, $J_3=0$.
In terms of fermions, the above Hamiltonian can be rewritten as
\begin{align}
H  =& 
- \frac{J_1}{N_f} \sum_{\<\v i\v j\>}  \psi^{n\bt}_{\v j}\psi^{n\al \dag}_{\v i}
 \psi^{m\al}_{\v i} \psi^{m\bt \dag}_{\v j} 
- \frac{J_2}{N_f} \sum_{\<\v i\v j\>}  \la^{a}_{\v j}\la^{a \dag}_{\v i}
 \la^{b}_{\v i} \la^{b \dag}_{\v j} 
\nonumber\\
& +\text{Const.} 
\end{align}
Using path integral, we can rewrite the above model as
\begin{align}
\label{pathint}
Z =& \int \cD(\psi^\dag)\cD(\psi)\cD(a_0) \cD(u) \cD(\chi) 
e^{i\int dt L} \nonumber\\
L =
& \psi^{n \dag}_{\v i}i[\prt_t+i a_0(\v i)] \psi^{n}_{\v i}
- \sum_{\<\v i\v j\>}  (\psi^{n \dag}_{\v i} u_{\v i\v j} \psi^{n}_{\v j}
+h.c.) 
\nonumber\\
& +\la^{a \dag}_{\v i}i[\prt_t+i \Tr a_0(\v i)] \la^{a}_{\v i}
- \sum_{\<\v i\v j\>}  (\la^{a \dag}_{\v i} \chi_{\v i\v j} \la^{a}_{\v j}
+h.c.) 
\nonumber\\
&- \frac{N_f}{J_1} \sum_{\<\v i\v j\>}  \Tr(u_{\v i\v j} u_{\v i\v j}^\dag)
- \frac{N_f}{J_2} \sum_{\<\v i\v j\>}  \chi_{\v i\v j} \chi_{\v i\v j}^\dag
\end{align}
where $(\psi^{n}_{\v i})^T =( \psi^{n,1}_{\v i}, \psi^{n,2}_{\v i},
\psi^{n,3}_{\v i}) $, and $a_0(\v i)$ and $u_{\v i\v j}$ are $3\times 3$
complex matrices that satisfy 
\begin{equation}
 u_{\v i\v j}^\dag =u_{\v j\v i},\ \ \
 a_0(\v i) = a^\dag_0(\v i)
\end{equation}
When $J_3\neq 0$, the Lagrangian may contain terms that mix $\chi_{\v i\v j}$
and $u_{\v i\v j}$:
\begin{align}
L =
& \psi^{n \dag}_{\v i}i[\prt_t+i a_0(\v i)] \psi^{n}_{\v i}
- \sum_{\<\v i\v j\>}  (\psi^{n \dag}_{\v i} u_{\v i\v j} \psi^{n}_{\v j}
+h.c.) 
\nonumber\\
& +\la^{a \dag}_{\v i}i[\prt_t+i \Tr a_0(\v i)] \la^{a}_{\v i}
- \sum_{\<\v i\v j\>}  (\la^{a \dag}_{\v i} \chi_{\v i\v j} \la^{a}_{\v j}
+h.c.) 
\nonumber\\
&- \frac{N_f}{J_1} \sum_{\<\v i\v j\>}  \Tr(u_{\v i\v j} u_{\v i\v j}^\dag)
- \frac{N_f}{J_2} \sum_{\<\v i\v j\>}  \chi_{\v i\v j} \chi_{\v i\v j}^\dag
\nonumber\\
& + C N_f \sum_{\<\v i\v j\>}[\chi_{\v i\v j}\det(u_{\v j\v i})+h.c.]
\end{align}
where $C$ is a $O(1)$ constant.
We note that the above Lagrangian describes a $U(1)\times SU(3)$ lattice
gauge theory coupled to fermions.

The field $a_0(\v i)$ in the Lagrangian is introduced to enforce the constraint
\begin{equation}
\psi^{n\al \dag}_{\v i}\psi^{n\bt}_{\v i}
- \psi^{n\bt}_{\v i} \psi^{n\al \dag}_{\v i}
+\la^{a \dag}_{\v i}\la^{a}_{\v i}\del^{\al\bt}
- \la^{a}_{\v i} \la^{a \dag}_{\v i} \del^{\al\bt}=0
\end{equation}
As in standard gauge theory, the above constraint really means a constraint
on physical states. \ie all physical states must satisfy
\begin{equation}
 \left(\la^{a \dag}_{\v i}\la^{a}_{\v i}\del^{\al\bt}+
\psi^{n\al \dag}_{\v i}\psi^{n\bt}_{\v i} -\del^{\al\bt}\frac{3}{2}N_f
\right)
\ket{\Phi_{phys}} =0
\end{equation}
The above is the needed constraint to obtain the Hilbert space of
our bosonic model.

Here we would like to stress that writing a bosonic model in terms of
gauge theory does not imply the existence physical gauge bosons at low energy.
Using projective construction, we can write any model in terms of a gauge
theory of any gauge group.\cite{BA8880,Wpcon} The existence of low energy
gauge fluctuations is a property of ground state. It has nothing to do with
how we write the Hamiltonian in terms of this or that gauge theory.

Certainly, if the ground state is known to have certain gauge fluctuations,
then writing Hamiltonian in term of a particular gauge theory that happen to
have the same gauge group will help us to derive the low energy effective
theory.  Even when we do not know the low energy gauge fluctuations in the
ground state, we can still try to write the Hamiltonian in a form that
contains certain  gauge theory and try to derive the low energy effective
gauge theory.  Most of the times, we find the gauge fluctuations in the low
energy effective theory is so strong that the gauge theory is in the confining
phase. This indicates that we have chosen a wrong form of Hamiltonian.
However, if we are lucky to choose the right form of Hamiltonian with right
gauge group, then the gauge fluctuations in the low energy effective theory
will be weak and the gauge fields $a_0$, $\chi_{\v i\v j}$ and $u_{\v i\v j}$
will be almost like classical fields. In this case, we can say that the ground
state of the Hamiltonian contains low energy gauge fluctuations described by
$a_0$, $\chi_{\v i\v j}$ and $u_{\v i\v j}$. In the following, we will show
that the $U(1)\times SU(3)$ fermion model \Eq{pathint} is the right form for
us to write the Hamiltonian \Eq{SUNfspin} of our bosonic model.

After integrating out the fermions, we obtain the following effective theory
for $a_0(\v i)$, $\chi_{\v i\v j}$ and $u_{\v i\v j}$
\begin{equation}
Z = \int \cD(a_0)\cD(u) e^{i\int dt N_f \t L_{eff}(u,a_0)} 
\end{equation}
where $\t L_{eff}$ does not depend on $N_f$. We see that, in the large $N_f$
limit, $\chi_{\v i\v j}$, $u_{\v i\v j}$ and $a_0$ indeed becomes classical
fields with weak fluctuations.

In the semi-classical limit, the ground state of the system is given by the
ansatz $(\bar \chi_{\v i\v j}, \bar u_{\v i\v j}, \bar a_0(\v i))$ that
minimize the energy $-\t L_{eff}$.  We will assume that such an ansatz have
$\pi$ flux on every plaquette and takes a form
\begin{align}
\label{U1a}
\bar \chi_{\v i,\v i+\hat{\v x}} =& -i\chi , &
\bar \chi_{\v i,\v i+\hat{\v y}} =& -i(-)^{i_x}\chi ,  \nonumber\\
\bar \chi_{\v i,\v i+\hat{\v z}} =& -i(-)^{i_x+i_y}\chi, \nonumber\\
\bar u_{\v i,\v i+\hat{\v x}} =& -iu , &
\bar u_{\v i,\v i+\hat{\v y}} =& -i(-)^{i_x}u ,  \nonumber\\
\bar u_{\v i,\v i+\hat{\v z}} =& -i(-)^{i_x+i_y}u, &
a_0(\v i) =& 0 . 
\end{align}
(If the $\pi$-flux ansatz does not minimize the energy, we can always modify
the Hamiltonian of our bosonic model to make the $\pi$-flux ansatz to have the
minimal energy.) Despite the $\v i$ dependence, the above ansatz actually
describe translation, rotation, parity, and charge conjugation symmetric
states. This is because the symmetry transformed ansatz, although not equal to
the original ansatz, is gauge equivalent to the original ansatz.

The mean-field Hamiltonian for fermions has a form 
\begin{equation}
 H = 
\sum_{\<\v i\v j\>}  (
\psi^{n \dag}_{\v i} \bar u_{\v i\v j} \psi^{n}_{\v j}
+\la^{a \dag}_{\v i} \bar \chi_{\v i\v j} \la^{a}_{\v j}
+h.c. )
\end{equation}
The fermion dispersion has two nodes at $\v k = 0$ and $\v k = (0,0,\pi)$.
Thus there are $2N_f\times 7$ massless four-component Dirac fermions in the
continuum limit.  They correspond to quarks and leptons of $2N_f$ different
families.  Each family contains six quarks (two flavors times three colors)
that carry $SU(3)$ colors and
charge $1/3$ for the $U(1)$ gauge field, and one lepton   that carry no
$SU(3)$ colors and charge $1$ for the $U(1)$ gauge field.

Including the collective fluctuations of the ansatz, 
the $U(1)\times SU(3)=U(3)$ fermion theory has a form
\begin{align}
\label{aUSU}
& L = 
\sum_{\v i} \psi^{n\dag}_{\v i} i (\prt_t + i a_0(\v i)) \psi^n_{\v j} 
+\sum_{\v i\v j}
\psi^{n\dag}_{\v i}\bar u_{\v i\v j}e^{ia_{\v i\v j}}\psi^n_{\v j} 
\nonumber\\
&+\sum_{\v i} \la^{a\dag}_{\v i} i (\prt_t + i \Tr a_0(\v i)) \la^a_{\v j} 
+\sum_{\v i\v j}
\la^{a\dag}_{\v i}\bar \chi_{\v i\v j}\det(e^{ia_{\v i\v j}})\la^a_{\v j} 
\end{align}
where $a_{\v i\v j}$ are $3\times 3$ hermitian matrices, describing $U(1)$ and
$SU(3)$ gauge fields. In the continuum limit, the above becomes 
\begin{align}
\label{U1SU3c}
 \cL = &
\bar \psi_{I, n} D_0 \ga^0 \psi_{I, n}
+v_f \bar \psi_{I, n} D_i \ga^i \psi_{I, n}
\nonumber\\
&+\bar \la_{I, a} D_0' \ga^0 \la_{I, a}
+v_f' \bar \la_{I, a} D_i' \ga^i \la_{I, a}
\end{align}
with 
$v_f\sim l_0J_1$, 
$v'_f\sim l_0J_2$, 
$D_\mu =\prt_\mu + i a_\mu$, $D'_\mu =\prt_\mu + i \Tr
a_\mu$, $I=1,2$, and $\ga^\mu$ are $4\times 4$ Dirac
matrices.\cite{BSK7643,ID89,Z92} $\la_{I,a}$ and $\psi_{I, n}$ are Dirac
fermion fields.  $\psi_{I, n}$ form a fundamental representation of color
$SU(3)$. 

If we integrate out $a_0$ and $a_{\v i\v j}$ in \Eq{aUSU} first, we will 
recover the bosonic model \Eq{SUNfspin}. 
If we integrating out the high energy fermions
first, the $U(1)\times SU(3)$ gauge field $a_\mu$ will acquire a dynamics.  We
obtain the following low energy effective theory in continuum limit
\begin{align}
\label{U1SU3cd}
 \cL = &
\bar \psi_{I, n} D_0 \ga^0 \psi_{I, n}
+v_f \bar \psi_{I, n} D_i \ga^i \psi_{I, n} \nonumber\\
&+\bar \la_{I, a} D_0' \ga^0 \la_{I, a}
+v_f' \bar \la_{I, a} D_i' \ga^i \la_{I, a} \nonumber\\
& + \frac{1}{\al_S}[\Tr F_{0i} F^{0i} 
+c_a^2 \Tr F_{ij} F^{ij} \big]
+ ...
\end{align}
where the velocity of the $U(3)$ gauge bosons is
$c_a\sim l_0 J_{1,2}$, and $...$ represents higher derivative terms and the
coupling constant $\al_S$ is of order $1/N_f$.  

In the large $N_f$ limit, fluctuations of the gauge fields are weak. The model
\Eq{U1SU3cd} describes a $U(1)\times SU(3)$ gauge theory coupled weakly to
$2N_f$ families of massless fermions.  Therefore, our bosonic model can
generate massless artificial quarks and artificial leptons that couple to
artificial light and artificial gluons. As discussed in \Ref{Wlight}, the PSG
of the ansatz \Eq{U1a} protects the masslessness of the  artificial quarks and
the artificial leptons. Our model has an emerging chiral symmetry.

\section{Conclusion}

In this paper, we studied a new class of ordered states - string-net condensed
states - in local bosonic models. The new kind of orders does not break any
symmetry and cannot be described by Landau's symmetry breaking theory. We show
that different string-net condensation can be characterized (and hopefully,
classified) by the projective symmetry in the hopping Hamiltonian for ends of
condensed strings. Similar to symmetry breaking states (or ``particle''
condensed states), string-net condensed states can also produce and protect
gapless excitations. However, unlike symmetry breaking states which can only
produce and protect gapless scaler bosons (or Nambu-Goldstone modes),
string-net condensed states can produce and protect gapless gauge bosons and
gapless fermions. It is amazing to see that gapless fermions can even appear
in local bosonic models.

Motivated by the above results, we propose the following locality principle:
\emph{The fundamental theory for our universe is a local bosonic model.} Using
several local bosonic models as examples, we try to argue that the locality
principle is not obviously wrong, if we assume that there is a string-net
condensation in our vacuum. The string-net condensation can naturally produce
and protect massless photons (as well as gluons) and (nearly) massless
electrons/quarks.  However, to really prove the string-net condensation in our
vacuum, we need to show that string-net condensation can generate chiral
fermions.  Also, the above  locality principle has not taken quantum gravity
into account. It may need to be generalized to include quantum gravity.  In
any case, we can say that we have a plausible understanding where light and
fermions come from. The existence of light and fermions is no longer
mysterious once we realize that they can come from local bosonic models via
string-net condensations.

The string-net condensation and the associated PSG also provide a new solution
to the chiral symmetry and the fermion mass problem in lattice QED and lattice
QCD. We show that the symmetry of a lattice bosonic model leads to PSG of the
hopping Hamiltonian for the ends of condensed strings.  If the ends of
condensed strings are fermions, then PSG can some times protect the
masslessness of the fermions, even though the chiral symmetry in the continuum
limit cannot be generalized to the lattice.  Thus PSG can lead to an emerging
chiral symmetry that protect massless Dirac fermions.

In this paper, we have been stressing that string-net condensation and the
associated PSG can protect the masslessness of fermions. However, most
fermions in nature do have masses, although very small comparing to the Planck
mass. One may wonder where those small masses come from. Here we would like to
point out that the PSG argument for masslessness only works for radiative
corrections. In other words, the fermions protected by string-net condensation
and PSG cannot gain any mass from additive radiative corrections caused by
high energy fluctuations. However, if the model has infrared divergence, then
infrared divergence can give the would-be-massless fermions some masses. The
acquired masses should have the scale of the infrared divergence. The 3+1D QED
model studied in this paper do not have any infrared divergence. Thus, the
artificial electrons in the model are exactly massless.  But in the bosonic
model discussed in section \ref{secQCD}, the $SU(3)$ gauge coupling $\al_S$
runs as
\begin{equation}
 \frac{d\al_S^{-1}}{d \ln(M^2)} = \frac{11-(2/3)(2N_f)}{4\pi}
\end{equation} 
where $M$ is the cut-off scale.  Thus when $N_f\leq 8$, $\al_S$ has a
logarithmic infrared divergence.  In general, for models with $U(1)$ and
$SU(3)$ gauge interactions and a right content of fermions, the $SU(3)$ gauge
interactions can have a weak logarithmic infrared divergence in
3+1D.\cite{W9985,H0049} This weak divergence could generate mass of order
$e^{-C/\al_S(M_P)}M_P$, where $M_P$ is the Planck mass or the GUT scale (the
cut-off scale of the lattice theory), $C=O(1)$ and $\al_S(M_P)$ is the
dimensionless gauge coupling constant at the Planck scale.  An
$C/\al_S(M_P)\sim 40$ can produce a desired separation between the Planck
mass/GUT scale  and the masses of observed fermions.  It is interesting to see
that, in order to use string-net condensation picture to explain the origin of
gauge bosons and nearly massless fermions, it is important to have a four
dimensional space-time.  When space-time has five or more dimensions, the
gauge-fermion interaction do not have any infrared divergence. In this case,
if a string-net condensation produces massless fermions, those fermions will
remain to be massless down to zero energy. In 2+1D, the gauge interaction
between massless fermions is so strong that one cannot have fermionic
quasiparticles at low energies.\cite{AN9021,RWspec,FT0103} It is amazing to
see that 3+1 is the only space-time dimension that the gauge bosons and
fermions produced by string-net condensation have weak enough interaction so
that they can be identified at low energies and, at same time, have strong
enough interaction to have a rich non-trivial structure at low energies.

This research is supported by NSF Grant No. DMR--01--23156
and by NSF-MRSEC Grant No. DMR--02--13282.

\end{document}